\documentstyle[preprint,prl,aps,graphicx]{revtex}

\def\beq{\begin{equation}}
\def\eeq{\end{equation}}
\def\beqa{\begin{eqnarray}}
\def\eeqa{\end{eqnarray}}

\def\za{\alpha}
\def\zb{\beta}
\def\lsim{\mathrel{\raise.3ex\hbox{$<$\kern-.75em\lower1ex\hbox{$\sim$}}} }
\def\gsim{\mathrel{\raise.3ex\hbox{$>$\kern-.75em\lower1ex\hbox{$\sim$}}} }

\begin{document}
\draft
\preprint{{\vbox{ \hbox{IPAS-HEP-k009}
\hbox{NCU-HEP-k005}\hbox{KUPT 02-04}\hbox{May 2002} 
\hbox{rev. Jul 2003}}}}

\title {A Detailed Analysis of One-loop Neutrino Masses from
the Generic Supersymmetric Standard Model}

\author{\bf Sin Kyu Kang}
\address{Institute for Basic Science, Korea University, Seoul 136-701, Korea}
\address{Graduate School of Science, Hiroshima University, 
            Higashi-Hiroshima 739-8526, Japan
\\E-mail: kang@theo.phys.sci.hiroshima-u.ac.jp}
\author{\bf Otto C. W. Kong}
\address{Department of Physics, National Central University, Chung-li, TAIWAN 32054\\
Institute of Physics, Academia Sinica, Nankang, Taipei, TAIWAN 11529
\\E-mail: otto@phy.ncu.edu.tw}
\maketitle

\begin{abstract}
{In the generic supersymmetric standard model which had no global
symmetry enforced by hand, lepton number violation is a natural
consequence. Supersymmetry, hence, can be considered the source of
experimentally demanded beyond standard model properties for the 
neutrinos. With an efficient formulation of the model, we perform a 
comprehensive detailed analysis of all (fermion-scalar) one-loop 
contributions to neutrino masses. }
\end{abstract} 	      

\newpage

\section{Introduction.}
Low-energy supersymmetry (SUSY) is the most popular candidate theory for
physics beyond the Standard Model (SM). The most extensively studied version
called the minimal supersymmetric standard model (MSSM) has an extra 
{\it ad hoc} discrete symmetry, called R parity, imposed on the Lagrangian.
It is defined in terms of baryon number, lepton number, and spin as,
explicitly, $R = (-1)^{3B+L+2S}$.
The consequence is that the accidental symmetries of baryon number and lepton
number in the SM are preserved, at the expense of making particles and
superparticles having a categorically different quantum number, R parity.
The latter is actually not the most effective discrete symmetry to control
superparticle mediated proton decay\cite{pd}, but is most restrictive in terms
of what is admitted in the Lagrangian, or the superpotential alone.

R parity also forbids neutrino masses in the supersymmetric SM. 
However, the recent data from the solar and atmospheric neutrino experiments
can be interpreted in terms of massive neutrino oscillations.
Thus, the strong experimental hints for the existence of (Majorana) neutrino
masses \cite{nu} is an indication of lepton number violation, hence suggestive of
R-parity violation. Giving up R parity, a tree-level neutrino mass can be 
generated through diagonalization of the neutrino-neutralino mass matrix.
At the 1-loop level, all three neutrinos will become massive. There is then
no need to introduce extra superfields beyond what is required by the
SM itself to describe neutrino phenomenology.
 
There is certainly no lack of studies on various RPV models
in the literature. However, such models typically involve strong assumptions 
on the form of R-parity violation.
In most cases, no clear statement on what motivates the assumptions taken
is explicitly given. In fact, there are quite some confusing, or even plainly
wrong, statements on the issues concerned. It is important to distinguish among 
the different RPV  "theories", and, especially, between such a theory and the
unique general supersymmetric standard model (GSSM) \cite{as12,as8}. The 
latter is the {\it complete} theory of SUSY without R-parity, one which admits 
all the RPV terms without {\it a priori} bias. In the GSSM, RPV terms
come in many different forms.
In  order not to miss any plausible RPV phenomenological features, it is 
important that all of the RPV parameters be taken into consideration.
A clear listing and discussion of all these is recently presented in 
Refs.\cite{as5}, under the framework of the single-VEV parametrization 
(SVP) \cite{ru1,ru2}. The latter, summarized below, is an optimal choice of 
flavor bases that helps to guarantee a consistent and unambiguous treatment of
all kind of admissible RPV terms with complete RPV effects on tree-level mass 
matrices for all states including scalars and fermions maintaining the simplest
structure. Following the formulation, we present here a complete list of
all the neutrino masses contributions up to 1-loop level.

A (Majorana) neutrino mass term violates SM lepton number by two units.
The experimental evidence for neutrino masses comes in through indications
of flavor oscillations, which requires mass mixings of the flavor states,
$\nu_e$, $\nu_\mu$, and $\nu_\tau$. Hence, we want neutrino mass terms
that have lepton flavor violation (LFV). The latter is a generic consequence
of R-parity violation.  To put it in another way, the GSSM in fact contains
many couplings that has one unit of LFV. Any combination of two of such
couplings may be able to give rise to a neutrino mass term. Since the expected
sub-eV neutrino masses are essentially the strongest source of upper bounds 
on such couplings\cite{rpv} up to the present moment, we have no way 
to tell which particular combinations 
of couplings do saturate the bounds and give a dominant contribution to a
neutrino mass term. In fact, each term also depends on a set of (R-parity
conserving) SUSY or MSSM parameters. We do not have much knowledge 
on the SUSY parameters beyond some lower bounds on a set of  related 
experimental parameters (mainly) from collider machines. In relation to the
neutrino mass contributions discussed here, the set of SUSY parameters
are typically taken as fixed by one generic SUSY mass scale. Changing 
the latter of course changes the actually neutrino masses resulted. More
importantly, it is not totally clear whether some phenomenological
hierarchy among values of the different SUSY parameters, may be together 
with some hierarchy among the values of the parameters with LFV, would not
give a picture on the relative importance of the various neutrino mass terms
different from what one may expect from such the kind of highly simplified 
analyses. Thus, it would be useful to have a complete list of such
neutrino mass terms without much {\it a priori} assumption involved.

Guided by theoretical prejudices or otherwise, many different pieces 
of such neutrino mass terms have been studied\cite{skk,ru6}
(see also Ref.\cite{DL} for a more updated list of references). More
recently, there are attempts to give the more complete story. In particular,
Ref.\cite{kias}, gives the general formulae for neutrino mass contributions up 
to the full 1-loop level.  However, the latter analysis is not formulated under 
the SVP and any detailed discussion is limited  to a scenario where the ``third
generation couplings dominate". Among the trilinear RPV couplings, this amounts
to admitting only non-zero $\lambda^{\!\prime}_{i\scriptscriptstyle 33}$'s and
$\lambda_{i\scriptscriptstyle 33}$'s, though all nonzero 
bilinear RPV are indeed included. The maximal mixing result from 
Super-Kamiokande may bring that  wisdom of ``third generation domination" under
question.  Refs.\cite{ru6} and \cite{as1}, for example, illustrate how no (family)
hierarchy, or even an anti-hierarchy, among the RPV couplings may be preferred.
More important to our perspective here, the study has assumption 
on the $B_i$ parameters and is interested in the numerical study of a specific 
high-energy scenario. Here, we aim at a more detailed analytical study
on the different pieces of contribution instead. With the help of a more
simple theoretical framework, the SVP, we follow the basic approach of
Ref.\cite{kias} and give a more transparent list of formulae, as well as 
pushing on to give much more detailed analytical results of each 
individual neutrino mass term.

The basic approach of Ref.\cite{kias} is to give each 1-loop neutrino mass
diagram in terms of effective couplings of the mass eigenstates of various
scalars and fermions running inside the loop, using a formula from the 
so-called ``effective mixing matrix" method\cite{neutra}.  
Details of all the admissible RPV contributions to
all the scalar, as well as fermion, mass terms under the SVP framework
are very manageable\cite{as4-7,as8}. The complete expressions,
together with useful perturbative diagonalization formulae for the interesting
elements of the mixing matrices are listed in Ref.\cite{as8}. We use below
exactly the same notation as presented in details in the latter reference,
which is taken as the background of the present presentation.  Our goal 
is to present the exact analytical expression for each neutrino mass 
term,  and the approximate dominating result from each term under very
mild assumption. The major part of the approximation is the perturbative
diagonalization formulae of the mass matrices, which are well founded 
on the smallness of the neutrino masses. The approximation also helps
to extract the major RPV parameter dependence of each mass term
and, hence, is an important target of the present study. 

There is actually a detailed analysis of all the neutrino mass terms
pretty much in the same spirit of present study published\cite{DL}.
The latter reference also essentially adopted the SVP framework.
However, mass insertion approximation is used to obtain the
results based on the use of MSSM states. Our approach here may be
a more direct and transparent alternative.  Having results from both
approaches also serving as a counter checking and helps to illustrate
more clearly some of the subtle points involved. Ref.\cite{DL} also
has some very different emphasis in their discussion. Hence, we consider
the present study necessary to complete the story of neutrino masses
in GSSM (or from R-parity violation). Moreover, our exact formulae, in terms
of mass eigenstates running inside the loop allow direct numerical calculations
of the neutrino mass results free from any approximation. We are also working 
on a detailed study of radiative neutrino decay within the model\cite{rd}, to
which the present paper also gives the necessary background. 

It should be emphasized here that it is not out intention to discuss scenarios
within the general model that could fit the experimental date. There being
such a large number of lepton number violating parameters within the
GSSM, phenomenologically viable scenarios will not be difficult to find.
The beauty of the GSSM in explaining the neutrino data is that the
parameters responsible will also give a rich collections of other 
experimental signals. More studies of various aspects of the model,
and constraints from various SUSY and LFV searches in the future
may give much better guideline for picking the real interesting scenarios.
The goal of the present study is to provide a useful better reference
for such efforts.

It should also be noted that we do not include here results of the gauge boson
loop contributions. Such contributions have been studied (see for example
article by Hempfling in Ref.\cite{skk}). They represent a small correction
to the tree-level results which could be absoluted into a renormalization of
the tree-level lepton flavor violating parameters, as also pointed out in
Ref.\cite{DL}. Hence, we focus only on the finite 1-loop contributions that
will give new structure to the neutrino mass matrix (same strategy was
adopted in Ref.\cite{DL}). Namely, we focus on the fermion-scalar loop.
Our formulation in terms of mass eigenstates for the fermions and scalars
inside the loop allow us to identify explicitly the Goldstone mode. Any
Goldstone mode contribution is taken out from the summation over mass
eigenstates running inside the loop. In fact, the Goldstone modes are
of course unphysical and calculation of their contribution gauge dependent.
The diagrams with the Goldstone modes only form gauge invariant sets
with the corresponding gauge boson and ghost diagrams added together.
If one wishes, one could take it that we are doing the fermion-scalar loop 
calculation in the unitary gauge. The gauge choice does not matter, as 
we are focusing on the sector of physical scalar bosons and fermions only. 
Including the Goldstone modes in our fermion-scalar loop calculations 
would rather be inconsistent.

In Sec.II below, we give a brief summary of the basic formulation of GSSM
used. Readers are referred to Ref.\cite{as8} for details. Sec.III then starts
on the neutrino mass discussion. While the tree-level neutrino-neutralino
mass matrix is quite well known, we present some of the details here for 
completeness. The presentation also sets the stage for the discussion of
the 1-loop contribution calculation. All the basics of the 1-loop analysis
is presented in the latter parts of the section. The next section discusses
some details of the results in the way outlined above. Some of the detailed
listing of individual terms are, however, left to the Appendices. In Sec.V,
we present a brief discussion on the application of the results to numerical
studies, while any detailed numerical studies will be leave for future 
publications. Finally, we conclude the paper with some remarks in Sec.VI. 

\section{Background of the GSSM}
Let us start with summarizing our formulation and notations here; readers are 
referred to Ref.\cite{as8} for more details.
The most general renormalizable superpotential with the spectrum of minimal
superfields containing all the SM states can be written as
\begin{equation}
W \!\! = \!\varepsilon_{ab}\left[ \mu_{\alpha}  \hat{H}_u^a \hat{L}_{\alpha}^b
+ h_{ik}^u \hat{Q}_i^a   \hat{H}_{u}^b \hat{U}_k^{\scriptscriptstyle C}
+ \lambda_{\alpha jk}^{\!\prime}  \hat{L}_{\alpha}^a \hat{Q}_j^b
\hat{D}_k^{\scriptscriptstyle C} +
\frac{1}{2}\, \lambda_{\alpha \beta k}  \hat{L}_{\alpha}^a
 \hat{L}_{\beta}^b \hat{E}_k^{\scriptscriptstyle C} \right] +
\frac{1}{2}\, \lambda_{ijk}^{\!\prime\prime}
\hat{U}_i^{\scriptscriptstyle C} \hat{D}_j^{\scriptscriptstyle C}
\hat{D}_k^{\scriptscriptstyle C} \;  ,
\end{equation}
where  $(a,b)$ are $SU(2)$ indices, $(i,j,k)$ are the usual family (flavor)
indices (going from $1$ to $3$). The 4 $\hat{L}_{\alpha}$'s, with the 
$(\za, \zb)$ indices as extended flavor indices going from $0$ to $3$, include the
usual leptonic doublets and the $H_d$ doublet. Four doublet superfields with
the same quantum number are needed for gauge anomaly cancelation. The
four are not {\it a priori} distinguishable. The rest of the superfield notations 
are obvious. Note that $\lambda$ is antisymmetric in the first two indices,
as required by the $SU(2)$  product rules, shown explicitly here with
$\varepsilon_{\scriptscriptstyle 12} =-\varepsilon_{\scriptscriptstyle 21}=1$.
Similarly, $\lambda^{\!\prime\prime}$ is antisymmetric in the last two indices
from $SU(3)_{\scriptscriptstyle C}$, though color contents are not shown here.

Doing phenomenological studies without specifying a choice of flavor bases is
ambiguous. It is like doing SM quark physics with 18 complex Yukawa couplings, 
instead of the 10 real physical parameters. As far as the SM itself is concerned, 
the extra 26 real parameters are simply redundant. There is simply no way to learn
about the 36 real parameters of Yukawa couplings for the quarks in some generic
flavor bases, so far as the SM is concerned. For instance, one can choose to
write the SM quark Yukawa couplings such that the down-quark Yukawa couplings
are diagonal, while the up-quark Yukawa coupling matrix is a product of (the
conjugate of) the CKM and the diagonal quark masses, and the leptonic Yukawa
couplings diagonal. Doing that has imposing no constraint or assumption onto the
model.  On the contrary, not fixing the flavor bases makes the connection between 
the parameters of the model and the phenomenological observables ambiguous.

In the case of the GSSM, the choice of flavor basis among the 4 $\hat{L}_\za$'s
is a particularly subtle issue, because of the fact that they are superfields the 
scalar parts of which could bear VEVs. A parametrization called the single-VEV
parametrization (SVP) has been advocated since Ref.\cite{ru1}.
The central idea is to pick a flavor basis such that only one among the 
$\hat{L}_\za$'s, designated as $\hat{L}_0$, bears a non-zero VEV. 
There is to say, the direction of the VEV, or the Higgs
field $H_d$, is singled out in the four dimensional vector space spanned by
the $\hat{L}_\za$'s. Explicitly, under the SVP, flavor bases are chosen such that :
1/  $\langle \hat{L}_i \rangle \equiv 0$, which implies
$\hat{L}_0 \equiv \hat{H}_d$;
2/  $y^{e}_{jk} (\equiv \lambda_{0jk} =-\lambda_{j0k})
=\frac{\sqrt{2}}{v_{\scriptscriptstyle 0}}\,{\rm diag}
\{m_{\scriptscriptstyle 1},
m_{\scriptscriptstyle 2},m_{\scriptscriptstyle 3}\}$;
3/ $y^{d}_{jk} (\equiv \lambda^{\!\prime}_{0jk})
= \frac{\sqrt{2}}{v_{\scriptscriptstyle 0}}\,{\rm diag}\{m_d,m_s,m_b\}$;
4/ $y^{u}_{ik}=\frac{\sqrt{2}}{v_{\scriptscriptstyle u}}\,
V_{\!\mbox{\tiny CKM}}^{\!\scriptscriptstyle T}\; {\rm diag}\{m_u,m_c,m_t\}$,
where $v_{\scriptscriptstyle 0}\equiv \sqrt{2} \, \langle \hat{L}_0 \rangle$
and $v_{\scriptscriptstyle u}\equiv \sqrt{2} \,
\langle \hat{H}_{u} \rangle$. 
A point to note is that the $m_i$'s above are, conceptually,  
not the charged lepton masses. The parametrization is optimal, apart from 
some minor redundancy in complex phases among the couplings. We 
simply assume all the admissible nonzero couplings within the SVP 
are generally complex. The big advantage of the SVP is that it gives 
the complete tree-level mass matrices of all the states (scalars 
and fermions) the simplest structure\cite{as8}.

Following our notation above, the soft SUSY breaking terms
of the Lagrangian, can be written as follow : 
\beqa
V_{\rm soft} &=& \epsilon_{\!\scriptscriptstyle ab}
  B_{\za} \,  H_{u}^a \tilde{L}_\za^b +
\epsilon_{\!\scriptscriptstyle ab} \left[ \,
A^{\!\scriptscriptstyle U}_{ij} \,
\tilde{Q}^a_i H_{u}^b \tilde{U}^{\scriptscriptstyle C}_j
+ A^{\!\scriptscriptstyle D}_{ij}
H_{d}^a \tilde{Q}^b_i \tilde{D}^{\scriptscriptstyle C}_j
+ A^{\!\scriptscriptstyle E}_{ij}
H_{d}^a \tilde{L}^b_i \tilde{E}^{\scriptscriptstyle C}_j   \,
\right] + {\rm h.c.}\nonumber \\
&+&
\epsilon_{\!\scriptscriptstyle ab}
\left[ \,  A^{\!\scriptscriptstyle \lambda^\prime}_{ijk}
\tilde{L}_i^a \tilde{Q}^b_j \tilde{D}^{\scriptscriptstyle C}_k
+ \frac{1}{2}\, A^{\!\scriptscriptstyle \lambda}_{ijk}
\tilde{L}_i^a \tilde{L}^b_j \tilde{E}^{\scriptscriptstyle C}_k
\right]
+ \frac{1}{2}\, A^{\!\scriptscriptstyle \lambda^{\prime\prime}}_{ijk}
\tilde{U}^{\scriptscriptstyle C}_i  \tilde{D}^{\scriptscriptstyle C}_j
\tilde{D}^{\scriptscriptstyle C}_k  + {\rm h.c.}
\nonumber \\
&+&
 \tilde{Q}^\dagger \tilde{m}_{\!\scriptscriptstyle {Q}}^2 \,\tilde{Q}
+\tilde{U}^{\dagger}
\tilde{m}_{\!\scriptscriptstyle {U}}^2 \, \tilde{U}
+\tilde{D}^{\dagger} \tilde{m}_{\!\scriptscriptstyle {D}}^2
\, \tilde{D}
+ \tilde{L}^\dagger \tilde{m}_{\!\scriptscriptstyle {L}}^2  \tilde{L}
  +\tilde{E}^{\dagger} \tilde{m}_{\!\scriptscriptstyle {E}}^2
\, \tilde{E}
+ \tilde{m}_{\!\scriptscriptstyle H_{\!\scriptscriptstyle u}}^2 \,
|H_{u}|^2
\nonumber \\
&& + \frac{M_{\!\scriptscriptstyle 1}}{2} \tilde{B}\tilde{B}
   + \frac{M_{\!\scriptscriptstyle 2}}{2} \tilde{W}\tilde{W}
   + \frac{M_{\!\scriptscriptstyle 3}}{2} \tilde{g}\tilde{g}
+ {\rm h.c.} \; ,
\label{soft}
\eeqa
where we have used ${H}_{d}$ in the place of the equivalent $\tilde{L}_0$
among the trilinear $A$-terms. Note that 
$\tilde{L}^\dagger \tilde{m}_{\!\scriptscriptstyle \tilde{L}}^2  \tilde{L}$,
unlike the other soft mass terms, is given by a $4\times 4$ matrix. 
Comparing with the MSSM case,
$\tilde{m}_{\!\scriptscriptstyle {L}_{00}}^2$ corresponds to
$\tilde{m}_{\!\scriptscriptstyle H_{\!\scriptscriptstyle d}}^2$ while
$\tilde{m}_{\!\scriptscriptstyle {L}_{0k}}^2$'s give new mass mixings.
The other notations are obvious. The writing of the soft terms in the above
form makes identification of the scalar mass terms straight forward. 
Recall that only the doublets ${H}_{u}$ and ${H}_{d}$ bear VEVs. The $A$-terms
in the second line of Eq.(\ref{soft}) hence do not contribute to scalar masses.

The SVP formulation also gives the complex equations  
\begin{equation} \label{tp3}
B_i \, \tan\!\beta 
=     \tilde{m}^2_{\!{\scriptscriptstyle L}_{\!{\scriptscriptstyle 0}i} }
+ \mu_{\scriptscriptstyle 0}^{*}\,\mu_i \; ,
\end{equation} 
reflecting the removed redundancy of  parameters in a generic $\hat{L}_\za$ 
flavor basis. They are nothing but the vanishing tadpole equations. They give 
consistence conditions among the involved parameters that should not be 
overlooked. The equations suggest that the $B_i$'s are expected to be 
suppressed, with respect to the  $B_{\scriptscriptstyle 0}$, as the $\mu_i$'s are,
with respect to $\mu_{\scriptscriptstyle 0}$. The 
$ \tilde{m}^2_{\!{\scriptscriptstyle L}_{\!{\scriptscriptstyle 0}i} }$ 
parameters in particular are missing in some of the relevant discussions
in the literature. From a different perspective, one may tend to think that
the parameters are similar to the $ \tilde{m}^2_{\!{\scriptscriptstyle L}_{\!ij} }$ 
parameters linked to soft flavor mixings. However, fixing 
 $ \tilde{m}^2_{\!{\scriptscriptstyle L}_{\!{\scriptscriptstyle 0}i} }$ 
in Eq.(\ref{tp3}) leads to definite relations between a $B_i$ and a $\mu_i$
term, which may not be satisfied. The parameters $B_i$, $\mu_i$,
and $ \tilde{m}^2_{\!{\scriptscriptstyle L}_{\!{\scriptscriptstyle 0}i} }$ 
are not independent free parameters, because of the fact that freely chosen values 
of the set of parameters in a top-down approach, in general, do not land the model
automatically into the single-VEV basis. The tadpole equations are incorporated
completely into the scalar mass matrices involved in our calculations\cite{as8}.

\section{Neutrino Masses}
The GSSM has seven neutral fermions corresponding to the three
neutrinos and four, heavy, neutralinos. The heavy states are supposed
to be mainly gauginos and higgsinos, but there is now admitted (RPV)
mixings among all seven neutral electroweak states. 
In the case of small $\mu_i$'s of interest, it is convenient to use an
approximate seesaw block diagonalization to extract the effective
neutrino mass matrix. Note that the
effective neutrino mass here is actually written in a basis which
is approximately the mass eigenstate basis of the charged leptons,
{\it i.e.}, the basis is roughly $(\nu_e, \nu_\mu, \nu_\tau)$.
The tree-level result is very well-known\cite{skk,ru6}.

\subsection{Getting the Neutrinos among the Neutral Fermions}
We use the basis
$(-i\tilde{B}, -i\tilde{W}, 
\tilde{h}_{\!\scriptscriptstyle u}^{\!\scriptscriptstyle 0^C}\,, 
\tilde{h}_{\!\scriptscriptstyle d}^{\!\scriptscriptstyle 0}\,, 
{l}_{\scriptscriptstyle 1}^{\scriptscriptstyle 0}\,,
{l}_{\scriptscriptstyle 2}^{\scriptscriptstyle 0}\,,
{l}_{\scriptscriptstyle 3}^{\scriptscriptstyle 0}\,) $
to write the $7 \times 7$ neutral fermion mass matrix 
${\cal M_{\!\scriptscriptstyle N}}$. 
Note that $\tilde{h}_{\!\scriptscriptstyle d}^{\!\scriptscriptstyle 0}
\equiv {l}_{\scriptscriptstyle 0}^{\scriptscriptstyle 0}$,
while $\tilde{h}_{\!\scriptscriptstyle u}^{\!\scriptscriptstyle 0^C}$
is the charge conjugate of the higgsino
$\tilde{h}_{\!\scriptscriptstyle u}^{\!\scriptscriptstyle 0}$.
For small $\mu_i$'s, we have
$(\, {l}_{\scriptscriptstyle 1}^{\scriptscriptstyle 0},
{l}_{\scriptscriptstyle 2}^{\scriptscriptstyle 0},
{l}_{\scriptscriptstyle 3}^{\scriptscriptstyle 0} \, ) 
\approx (\nu_{\scriptscriptstyle e},\nu_{\scriptscriptstyle \mu}, 
\nu_{\scriptscriptstyle \tau})$\cite{as8}.
The symmetric, but generally complex, matrix can be diagonalized 
by using unitary matrix {\boldmath $X$} such that
\beq
\mbox{\boldmath $X$}^{\!\scriptscriptstyle  T} 
{\cal M_{\scriptscriptstyle N}}\mbox{\boldmath $X$} =
\mbox{diag} \{ {M}_{\!\scriptscriptstyle \chi^0_{n}} \} \; .
\eeq
Again, the first part of the mass eigenvalues, 
${M}_{\!\scriptscriptstyle \chi^0_{n}} $ for $n = 1$--$4$ here, gives
the heavy states, {\it i.e.} neutralinos. The last part, 
${M}_{\!\scriptscriptstyle \chi^0_{n}} $ for $n = 5$--$7$, 
hence gives the physical neutrino masses.

The mass matrix ${\cal M_{\!\scriptscriptstyle N}}$ can be written in the form 
of block submatrices:
\begin{equation} \label{mnu}
{\cal M_{\!\scriptscriptstyle N}} = \left( \begin{array}{cc}
              {\cal M}_n & \xi^{\!\scriptscriptstyle T} \\
              \xi & m_\nu^o \end{array}  \right ) \;,
\end{equation}
where ${\cal{M}}_n$ is the upper-left $4\times 4$ neutralino mass matrix, 
$\xi$ is the $3\times 4$ block, and $m_\nu^o$ is the lower-right 
$3\times 3$ neutrino block in the $7\times 7$ matrix. 
In the interest of small neutrino masses, a perturbative (seesaw) 
block diagonalization can be applied. Explicitly, the diagonalizing 
matrix can be written approximately as
\beq \label{BZm}
\mbox{\boldmath $Z$}\simeq
\left( \begin{array}{cc}
I_{\!\scriptscriptstyle 4 \times 4} & ({\cal M}_n^{\mbox{-}1} \, 
\xi^{\!\scriptscriptstyle  T}) \\
-({\cal M}_n^{\mbox{-}1} \,  \xi^{\!\scriptscriptstyle  T})^\dag  
&  I_{\!\scriptscriptstyle 3 \times 3}
\end{array} \right) \; 
\eeq 
The tree-level effective neutrino mass matrix may  then be obtained as 
\beqa \label{ssA}
(m_\nu)  \simeq 
&& 
- ({\cal M}_n^{\mbox{-}1} \,  \xi^{\!\scriptscriptstyle  T})
^{\scriptscriptstyle T} \, {\cal M}_n \, ({\cal M}_n^{\mbox{-}1} \, 
\xi^{\!\scriptscriptstyle  T})
= - \, \xi \, {\cal M}_n^{-1} \, \xi^{\!\scriptscriptstyle T} \,
\nonumber \\
\simeq &&   \frac{  M_{\!\scriptscriptstyle Z}^2 \, \cos^2\!\!\beta \,
 (M_{\!\scriptscriptstyle 1} \, \cos^2\!\theta_{\!\scriptscriptstyle W} \;
 + M_{\!\scriptscriptstyle 2} \, \sin^2\!\theta_{\!\scriptscriptstyle W} )}
{ \det({\cal M}_n) }  \;\;
(\, \mu_i \, \mu_j \,) \;\;
\eeqa
where
\begin{equation} 
\det({\cal M}_n) = \mu_{\scriptscriptstyle 0} \,
\left[-\mu_{\scriptscriptstyle 0} \,
M_{\!\scriptscriptstyle 1}\, M_{\!\scriptscriptstyle 2} \,
+ M_{\!\scriptscriptstyle Z}^2 \, \sin\!2\beta \,
 ( M_{\!\scriptscriptstyle 1} \, \cos^2\!\theta_{\!\scriptscriptstyle W} 
+ M_{\!\scriptscriptstyle 2} \, \sin^2\!\theta_{\!\scriptscriptstyle W} )
\right] 
\end{equation}
is equivalent in expression to the determinant of the MSSM neutralino mass matrix. 

It is obvious that the $3 \times 3$ matrix $(\,\mu_i \, \mu_j\,)$ has only 
one nonzero eigenvalue given by
\beq
\mu_{\scriptscriptstyle 5}^2 = |\mu_{\scriptscriptstyle 1}|^2 + 
 |\mu_{\scriptscriptstyle 2}|^2 + |\mu_{\scriptscriptstyle 3}|^2 \; .
\eeq
We can define
\beq
R_{\scriptscriptstyle 5} =  \left( \begin{array}{ccc}
\frac{\mu_{\scriptscriptstyle 1}^*}{\mu_{\scriptscriptstyle 5}}
	& 0 & \frac{\sqrt{|\mu_{\scriptscriptstyle 2}|^2 + 
|\mu_{\scriptscriptstyle 3}|^2}}{\mu_{\scriptscriptstyle 5}}
 \\
\frac{\mu_{\scriptscriptstyle 2}^*}{\mu_{\scriptscriptstyle 5}}
 	& \frac{\mu_{\scriptscriptstyle 3}}{\sqrt{|\mu_{\scriptscriptstyle 2}|^2
	   + |\mu_{\scriptscriptstyle 3}|^2}}
	& - \frac{\mu_{\scriptscriptstyle 1} \, \mu_{\scriptscriptstyle 2}^*}
{\mu_{\scriptscriptstyle 5} \, \sqrt{|\mu_{\scriptscriptstyle 2}|^2 
           + |\mu_{\scriptscriptstyle 3}|^2}}
 \\
\frac{\mu_{\scriptscriptstyle 3}^*}{\mu_{\scriptscriptstyle 5}}
	& -\frac{\mu_{\scriptscriptstyle 2}}{\sqrt{|\mu_{\scriptscriptstyle 2}|^2
	   + |\mu_{\scriptscriptstyle 3}|^2}}
	& - \frac{\mu_{\scriptscriptstyle 1} \, \mu_{\scriptscriptstyle 3}^*}
{\mu_{\scriptscriptstyle 5} \, \sqrt{|\mu_{\scriptscriptstyle 2}|^2
           + |\mu_{\scriptscriptstyle 3}|^2}}
\end{array} \right) \; .
\eeq
Then, we have  $R_{\scriptscriptstyle 5}^{\scriptscriptstyle  T} \, 
(\,\mu_i \, \mu_j\,) \, R_{\scriptscriptstyle 5} = \rm{diag}\{\,
\mu_{\scriptscriptstyle 5}^2, 0, 0 \, \}$. Here, $\mu_{\scriptscriptstyle 5}$ and
$\sqrt{|\mu_{\scriptscriptstyle 2}|^2 + |\mu_{\scriptscriptstyle 3}|^2}$ are taken
as real and positive. With this result, we can write the overall diagonalizing 
matrix {\boldmath $X$} in the form
\beq
\mbox{\boldmath $X$} \simeq
\left( \begin{array}{cc}
I_{\!\scriptscriptstyle 4 \times 4} &  ({\cal M}_n^{\mbox{-}1} \, 
\xi^{\!\scriptscriptstyle  T})\\
- ({\cal M}_n^{\mbox{-}1} \, 
\xi^{\!\scriptscriptstyle  T})^\dag  &  I_{\!\scriptscriptstyle 3 \times 3}
\end{array} \right) \,
\left( \begin{array}{cc}
R_{\scriptscriptstyle n} &  \quad  0_{\scriptscriptstyle 4 \times 3} \\
0_{\scriptscriptstyle 3 \times 4}  &    \quad e^{i\zeta} \,
R_{\scriptscriptstyle 5}
\end{array} \right)
= \left( \begin{array}{cc}
R_{\scriptscriptstyle n} &   \quad e^{i\zeta} \,  ({\cal M}_n^{\mbox{-}1} \, 
\xi^{\!\scriptscriptstyle  T}) \, 
R_{\scriptscriptstyle 5}\\
- ({\cal M}_n^{\mbox{-}1} \, 
\xi^{\!\scriptscriptstyle  T})^\dag \, 
R_{\scriptscriptstyle n} &     \quad e^{i\zeta} \, 
R_{\scriptscriptstyle 5}
\end{array} \right) \; ,
\eeq
where $R_{\scriptscriptstyle n}$ is a $4 \times 4$ matrix with elements all
expected to be of order 1, basically the diagonalizing matrix for the ${\cal M}_n$
block and $e^{i\zeta}$ is a constant phase factor put in to absorb the overall
phase in the constant factor in the expression of Eq.(\ref{ssA}) so that the
resulted neutrino mass eigenvalue would be real and positive. The matrix 
{\boldmath $X$} contains the important information of the gaugino and Higgsino 
contents of the physical neutrinos. This is given by the mixing elements in the 
off-diagonal blocks. The  {\boldmath $Z$} matrix in itself gives similar information
for the effective SM neutrinos (flavor states). The latter matrix may be more useful
in the analysis of neutrino phenomenology.

\subsection{Approach to 1-loop Neutrino Masses Calculations}
Following Ref.\cite{kias}, we use the 1-loop (renormalized) mass formula from
the ``effective mixing matrix" approach, giving a fermion mass matrix as
\beq \label{mpole}
{\cal M_{\!\scriptscriptstyle N}}^{\mbox{\tiny (1)}}(p^2) = {\cal M}(Q) + \Pi(p^2) -
{1\over 2}\, \left[\, {\cal M}(Q) \, \Sigma(p^2) + \Sigma(p^2) \, {\cal M}(Q)
\, \right] \; ,
\eeq
Note that ${\cal M}(Q)$ is the $\overline{DR}$ renormalized tree-level
mass (matrix), while $\Pi$ and $\Sigma$ the contributions from one-loop self-energy diagrams with and without chirality flip.  We have
\beq \label{emmf}
 {\cal M_{\!\scriptscriptstyle N}}^{\mbox{\tiny (1)}}(p^2) = 
\pmatrix{ {\cal M}_n & \xi^{\!\scriptscriptstyle  T} \cr  \xi & 0 \cr }
+ \pmatrix{ \delta\!{\cal M}_n & \delta\xi^{\!\scriptscriptstyle  T} 
\cr  \delta\xi & \delta(m_\nu^o) \cr }\!\!(p^2) \; ,
\eeq
where
\begin{eqnarray} 
\delta\!{\cal M}_n(p^2) &=&  \Pi_n(p^2)
- \frac{1}{2} \left[ \,  {\cal M}_n \, \Sigma_n(p^2)
 + \Sigma_n(p^2)  \, {\cal M}_n \,  \right] \; ,
 \nonumber\\
\delta\xi(p^2) &=& \Pi_\xi(p^2)
- \frac{1}{2} \left[ \, \Sigma_\nu(p^2) \, \xi +     \xi \, \Sigma_n(p^2)
 + \Sigma_\xi(p^2) \, {\cal M}_n \, \right]  \;  ,
\nonumber\\
\delta m_\nu^o(p^2) &=& \Pi_\nu(p^2) 
- \frac{1}{2} \left[ \,  \xi  \, \Sigma_\xi^{\!\scriptscriptstyle  T}(p^2)
+ \Sigma_\xi(p^2) \, \xi^{\!\scriptscriptstyle  T}  \,  \right]   \; ,
\eeqa
with the explicit renormalization scale ($Q$-)dependence of the tree-level 
parameters dropped. Seesaw diagonalization of 
${\cal M_{\!\scriptscriptstyle N}}^{\mbox{\tiny (1)}}$
yields the 1-loop result 
\beqa
(m_\nu)^{\mbox{\tiny (1)}} &\simeq&
- \xi \, {\cal M}_n^{\mbox{-}1} \, \xi^{\!\scriptscriptstyle  T}
+ \delta(m_\nu^o) 
- \delta\xi \, {\cal M}_n^{\mbox{-}1} \, \xi^{\!\scriptscriptstyle  T}
- \xi \, {\cal M}_n^{\mbox{-}1} \, \delta\xi^{\!\scriptscriptstyle  T}
+ \xi \, {\cal M}_n^{\mbox{-}1} \, \delta\!{\cal M}_n \,
{\cal M}_n^{\mbox{-}1} \, \xi^{\!\scriptscriptstyle  T}
\nonumber\\
&=& - \xi \, {\cal M}_n^{\mbox{-}1} \, \xi^{\!\scriptscriptstyle  T}
+ \Pi_\nu + \Pi_\xi \,  {\cal M}_n^{\mbox{-}1} \,  \xi^{\!\scriptscriptstyle  T}
+ \xi \, {\cal M}_n^{\mbox{-}1} \, \Pi_\xi^{\!\scriptscriptstyle  T}
\nonumber\\ 
&+& {1\over 2} \, \Sigma_\nu \, \xi \,  {\cal M}_n^{\mbox{-}1} \,  \xi^{\!\scriptscriptstyle  T}
+ {1\over 2}  \, \xi \,  {\cal M}_n^{\mbox{-}1} \,  \xi^{\!\scriptscriptstyle  T}
\, \Sigma_{\nu}^{\!\scriptscriptstyle  T} 
+ \xi \, {\cal M}_n^{\mbox{-}1} \, \Pi_n \,
{\cal M}_n^{\mbox{-}1} \, \xi^{\!\scriptscriptstyle  T} \; ,
\label{m1loop}
\eeqa
where we have dropped the $p^2$ dependence. As discussed below,
the $p^2$ should be taken as at the scale of the mass
${\cal M_{\!\scriptscriptstyle N}}$ itself. Hence, in the application here
to calculate the neutrino masses, the  $p^2$ in
$(m_\nu)^{\mbox{\tiny (1)}} (p^2)$ may be taken as  practically zero.
An important point to note here is that the $\Sigma_\xi$ and $\Sigma_n$
terms all cancel out and disappear from our final result for
$(m_\nu)^{\mbox{\tiny (1)}}$. We refer the reader to Refs.\cite{kias,neutra} for 
further discussion on the merits of the approach and references to related works. 

At this point, some remarks on the renormalization issue are in order. The issue
has been well-addressed in the papers by Hempfling and Hirsch 
{\it et.al.}\cite{skk} on calculations starting from tree-level mass
eigenstates. The analog formula to Eq.(\ref{mpole}) above is, under the 
$\overline{DR}$ scheme, 
\beq
{\cal M}^{\mbox{\tiny pole}} = {\cal M}^{\overline{DR}}(Q) +
\Delta {\cal M}(p,Q)
\eeq
where the 1-loop correction part is given as 
\beq \label{DM}
\Delta {\cal M}(p,Q) = [\Delta {\cal M}(p)]_{\Delta =0} \;,
\eeq
{\it i.e.} the two-point functions involved are calculated by subtracting the
term proportional to the regulator
$\Delta \equiv\frac{2}{4-d}-\gamma_{\!\scriptscriptstyle  E} + ln\,4\pi$
of dimensional reduction. There is some ambiguity in the choice of $Q$ in
the evaluation of the off-diagonal two-point functions. As pointed out in 
Hempfling's paper, the effect of the ambiguity is of higher order. In the 
``effective mixing matrix" approach\cite{neutra}, the equation is casted
in the electroweak state basis instead to arrive at Eq.(\ref{mpole}), which
upon seesaw block-diagonalization yields Eq.(\ref{m1loop}). Now, $p^2$ is
practically zero, as we are calculating only diagrams with neutrinos on the
external legs of the two-point functions. The rest are only tree-level
mass matrix entries (to ${\cal M_{\!\scriptscriptstyle N}}$) coming into
the formula as mixing matirx elements between the neutralino and neutrino
blocks. The result for $(m_\nu)^{\mbox{\tiny (1)}}$ is however 
$Q$-dependent. Apart from the $Q$-dependence in the former set of parameters,
there is also the $Q$-dependence coming up from the calculation of the
two-point functions under the $\overline{DR}$ scheme as in Eq.(\ref{DM}).
Furthermore, there are the full set of couplings involved in such 
calculations, which should be taken as running couplings at the scale $Q$.
In the straight formal sense, the pole mass formula gives result
${\cal M}^{\mbox{\tiny pole}}$ that is $Q$-independent. However, in the
application to obtain
 ${\cal M_{\!\scriptscriptstyle N}}^{\mbox{\tiny (1)}}(p^2)$ and
its subsequent use in any explicit calculations,  some residual $Q$-dependence 
is difficult to avoid. 

Since we are interested in radiative neutrino mass generation from 
superparticles, we may take $Q$ as $M_{\mbox{\tiny SUSY}}$, or roughly 
the electroweak scale. Below the scale, the superparticles decouple and 
the neutrino mass terms can only be expressed
in terms of five-dimensional operators of the SM. Strictly speaking, one get 
the correct pole mass for the neutrinos only by running the operators to
the neutrino mass scale through the corresponding renormalization group
equations. However, such effects are minimal\cite{nrge}. Apart from yielding
the neutrino mass matrix in the more interesting flavor basis, Eq.(\ref{m1loop})
also avoids the superficial singularity reflecting the arbitrariness in the 
diagonalization of the mass matrix with degenerate massless neutrinos at 
tree-level. Furthering, the MNS (neutrino mixing) matrix obtained from the
diagonalization of $(m_\nu)^{\mbox{\tiny (1)}}$ maintains a full
unitary matrix.

The full $7\times 7$ neutral fermion mass matrix has four heavy and three very 
light mass eigenvalues, corresponding to the neutralinos and neutrinos; and
we are essentially only interested in the neutrino states. For a general  mass 
calculation at 1-loop, we must choose a renormalization prescription for each of 
the tree-level parameters appearing in the full mass matrix 
${\cal M_{\!\scriptscriptstyle N}}$, and will have to worry about the 
renormalization scale dependence issues of such parameters. The ``effective 
mixing matrix"  formula [{\it cf.} Eq.(\ref{m1loop})] avoid the complication as the 
neutrino mass results depend only on $p^2$ explicitly, which is practical zero,
as pointed out in Ref.\cite{kias}.  The parameters involved in the neutrino
mass generation are then taken as running parameter at the scale of interest. Renormalization scale dependence comes in only through the 
$\Sigma_\nu$ part (readers of refer to Ref.\cite{neutra} for more details) 
and the effect is small.  The $\Sigma_\nu$ part itself is mostly not very
important, as can be easily seen in Eq.(\ref{m1loop}).

Our neutrino mass formula [Eq.(\ref{m1loop})] calls for a seesaw type block 
diagonalization of the mass matrix ${\cal M_{\!\scriptscriptstyle N}}$ up to 1-loop 
order. The diagonalizing transformation corresponds to the matrix {\boldmath $Z$}
of expression  (\ref{BZm}). The tree-level contribution, given by the first 
term in the formula, is obviously seesaw suppressed (by the neutralino mass 
scale). The second term $\Pi_\nu$ gives the direct 1-loop contributions. 
However, there are parts of $\Pi_\nu$ that involved other suppression
beyond the loop factor. A typical example is the pure gaugino loop, or
GH-loop\cite{GH}, diagram contribution which can be interpreted as 
requiring seesaw induced Majorana-like ``sneutrino" mass to give a 
nonvanishing result\cite{as5}. They may be called pseudo-direct 1-loop 
contributions. For the rest of the terms in Eq.(\ref{m1loop}), are indirect
1-loop contributions, which has part of the basic seesaw suppression going 
along. These include results from 1-loop diagrams contributing to the off-diagonal 
blocks of the ${\cal M_{\!\scriptscriptstyle N}}$ matrix, from $\Sigma_\nu$
diagrams, as well as from diagrams contributing to the diagonal block
${\cal M}_n$. The last one, given by the last term in the formula, 
gives no interesting features. It can be absorbed, for instance, into the
tree-level result (first term) by replacing ${\cal M}_n$ there with the
1-loop corrected result. And, from the related calculations within MSSM,
we know that the correction is about $6\%$\cite{neutra}. In fact, the
flavor conserving part of the contributions involving $\Sigma_\nu$ is
similarly uninteresting. However, the part of the latter with LFV may 
be  of interest.

To calculate explicitly the various neutrino mass contributions using the
above formula, we need to have the effective couplings of the electroweak
state neutral fermions to possible scalar and fermion mass eigenstates
running in the quantum loop. The neutral fermion themselves,
together with the nine neutral scalars of the model, give a class of neutral 
loop contributions. Obviously, the loop with the neutralino states dominates 
here. The effective couplings, to be given below, involve diagonalizing
matrix elements of the states contributing to the states running inside
the loop. For the fermion part, it is the {\boldmath $X$} matrix discussed
above. Similar perturbative diagonalization expressions for all the other
matrices, those for the charged fermion, charged scalar, down-squarks,
as well as the neutral scalar sector are discussed in details in Ref.\cite{as8}.
We refrain from repeating the long list of such formulae in this paper.
Most parts of the notation used, as will appear below, are quite easy to
appreciate. Readers interested in checking any details on the derivations
of the results, however, would need to use Ref.\cite{as8} extensively.

\subsection{ Neutral Loop Contributions}
For the neutral loop contributions, we start with the effective interaction
for the external ${l}_i^{\scriptscriptstyle 0}$'s with internal mass eigenstates, 
\begin{equation}
\label{intN}
{\cal L} = {g_{\scriptscriptstyle 2}}  \,
\overline{\Psi}({l}_i^{\scriptscriptstyle 0}) \;
\left[ {\cal N}^{\scriptscriptstyle L}_{\!\scriptscriptstyle inm}\, 
{1 - \gamma_{\scriptscriptstyle 5} \over 2} + 
{\cal N}^{\scriptscriptstyle R}_{\!\scriptscriptstyle inm} \, 
{1 + \gamma_{\scriptscriptstyle 5} \over 2} \right ] \,  
 \Psi({\chi}^{\scriptscriptstyle 0}_n) \; \phi_{m}^{\scriptscriptstyle 0} \;
+ \mbox{h.c.} \; ,
\end{equation}
where ${1\over 2}(1 \mp \gamma_{\scriptscriptstyle 5} )$ are the $L$- and
$R$-handed projections. We have 
\beq
{\cal N}^{\scriptscriptstyle R}_{\!\scriptscriptstyle inm}
= \frac{1}{{2}} \,
[ \tan\!\theta_{\!\scriptscriptstyle W}
\mbox{\boldmath $X$}_{\!\!1n}^{*} - \mbox{\boldmath $X$}_{\!\!2n}^{*} ] \;
  [ {\cal D}^{s}_{\!(i+2)m} + i \, {\cal D}^s_{\!(i+7)m} ] \; ,
\eeq
and ${\cal N}^{\scriptscriptstyle L}_{\!\scriptscriptstyle inm}
= {\cal N}^{\scriptscriptstyle R^*}_{\!\scriptscriptstyle inm}$. 
The direct 1-loop contributions is given by
$\!\!$\footnote{
Here, we have all fermions involved being Majorana fermions. We compose the 4-spinor $\Psi$ by 
$\Psi = \left(\begin{array}{c}  \psi_{\scriptscriptstyle\! L} \\ 
\psi_{\scriptscriptstyle\! R}  \end{array} \right)$    where we have
$ \psi_{\!\scriptscriptstyle R} = -i\,\sigma_{\!\scriptscriptstyle 2}  \psi_{\!\scriptscriptstyle L}^*$.
A mass term has $\overline{\Psi}^{\prime}{\Psi}= \psi_{\!\scriptscriptstyle L}^{\prime\dag}
 \psi_{\!\scriptscriptstyle R} +  \psi_{\!\scriptscriptstyle R}^{\prime\dag}  \psi_{\!\scriptscriptstyle L}$. 
For instance, the $\psi_{\!\scriptscriptstyle R}^{\prime\dag}  \psi_{\!\scriptscriptstyle L}$ 
part can be written as $\psi_{\!\scriptscriptstyle L}^{\prime\scriptscriptstyle T}\, i \, 
\sigma_{\!\scriptscriptstyle 2} \, \psi_{\!\scriptscriptstyle L}$.  The 
${\cal N}^{\scriptscriptstyle R^*}_{\!\scriptscriptstyle inm}$ vertex brings in a 
$\psi_{\!\scriptscriptstyle L_i}$  while a matching 
${\cal N}^{\scriptscriptstyle L}_{\!\scriptscriptstyle jnm}$ vertex brings in a 
$\psi_{\!\scriptscriptstyle R_j}^{\dag}=\psi_{\!\scriptscriptstyle L_j}^{\scriptscriptstyle T}\, i \, 
\sigma_{\!\scriptscriptstyle 2}$. With the proper handling of the fermion wavefunction, 
$\Pi_{\nu_{ij}}^{\scriptscriptstyle N}$ has contributions proportional to  
${\cal N}^{\scriptscriptstyle R^*}_{\!\scriptscriptstyle inm} \,
{\cal N}^{\scriptscriptstyle L}_{\!\scriptscriptstyle jnm}$ which is equivalent to 
the ${\cal N}_{\!\scriptscriptstyle inm}^{\scriptscriptstyle R^*} \,
 {\cal N}_{\!\scriptscriptstyle jnm}^{\scriptscriptstyle R^*} $ used here.
Just for the  ${l}_i^{\scriptscriptstyle 0}$ states, the use of 
${\cal N}_{\!\scriptscriptstyle inm}^{\scriptscriptstyle R}$ can only be seen as
the intrinsic left-handed nature of the states. However, the explicit use of their
charge conjugate to compose the 4-spinors and derive the effective couplings is 
necessary to complete the formulation, say in the case of the 
${\cal C}^{\scriptscriptstyle L}_{\!\scriptscriptstyle inm}$
for the charged loop discussed below. 
}
\beq
\Pi_{\nu_{ij}}^{\scriptscriptstyle N} = -
{\alpha_{\mbox{\tiny em}} \over 8 \pi \, 
\sin\!^2\theta_{\!\scriptscriptstyle W}} \;
 {\cal N}_{\!\scriptscriptstyle inm}^{\scriptscriptstyle R^*} \,
 {\cal N}_{\!\scriptscriptstyle jnm}^{\scriptscriptstyle R^*} \;
{M}_{\!\scriptscriptstyle \chi^0_{n}} \, 
{\cal B}_{0}\!\!\left(p^2,{M}_{\!\scriptscriptstyle \chi^0_{n}}^2,
 {M_{\!\scriptscriptstyle S_{m}}^2} \right) \; ,
\eeq
 where the loop function ${\cal B}_{0}$ is defined in the limit of 
$p^2 \to 0 $ by  
\begin{equation}
{\cal B}_{0}(p^2,m_1^2,m_2^2)=-{m_2^2 \over m_1^2-m_2^2} \ln {m_1^2 \over m_2^2}
        -\ln{m_1^2 \over Q^2} +1  
\end{equation}
As will be shown explicitly below, this result is the gauge loop contribution 
first discussed in Ref.\cite{GH}. Note that $\mbox{\boldmath $X$}$ is the
matrix that diagonalizes the seven neutral fermions, as discussed explicitly
above. Among the seven fermion (tree-level) mass eigenstates denoted by the
sum over $n$ here, contributions from the $n=5$-$7$ states are certainly
negligible. The sum of $m$ runs through the nine physical neutral scalar states.
The states, together with the unphysical Goldstone mode, are obtained from
the $10\times 10$ neutral scalar mass-squared matrix to be diagonalized by 
${\cal D}^s$. We refer readers to Ref.\cite{as8} for details on the scalar sector. 
The set of coupling vertices may also be combined to give contributions to the
self energy function $\Sigma_{\nu}$. We have
\beq
\Sigma_{\nu_{ij}}^{\scriptscriptstyle N} = -
{\alpha_{\mbox{\tiny em}} \over 8 \pi \, 
\sin\!^2\theta_{\!\scriptscriptstyle W}} \;
 {\cal N}_{\!\scriptscriptstyle inm}^{\scriptscriptstyle R^*} \,
 {\cal N}_{\!\scriptscriptstyle jnm}^{\scriptscriptstyle R} \;
{\cal  B}_{1}\!\!\left(p^2,{M}_{\!\scriptscriptstyle \chi^0_{n}}^2,
 {M_{\!\scriptscriptstyle S_{m}}^2} \right) \; ,
\eeq
where the loop function ${\cal B}_{1}$ is defined by in the limit of
$ p^2 \to 0 $ by
\begin{equation}
{\cal B}_{1}(p^2,m_1^2,m_2^2)=
   {1 \over 2} \left[ 1 - \ln {m_2^2 \over Q^2} \, \;
  -\left( {m_2^1 \over m_1^2-m_2^2} \right)^2 \ln {m_1^2 \over m_2^2} \, \;
  +{1 \over 2}\left( {m_1^2 +m^2_2 \over m_1^2 - m_2^2} \right) \right] \; .
\end{equation}

For the indirect 1-loop contributions, we need
\beqa
{\cal N}^{\scriptscriptstyle R}_{\!\scriptscriptstyle 0nm} 
&=& \frac{1}{{2}} \,       
[ \tan\!\theta_{\!\scriptscriptstyle W} 
\mbox{\boldmath $X$}_{\!\!1n}^{*} - \mbox{\boldmath $X$}_{\!\!2n}^{*} ] \;
  [ {\cal D}^{s}_{\!2m} + i \, {\cal D}^s_{\!7m} ]\; ,
\\
{\cal N}^{\scriptscriptstyle R}_{\!\scriptscriptstyle\tilde{h}\!nm} 
&=& - \frac{1}{{2}} \,       
[ \tan\!\theta_{\!\scriptscriptstyle W} 
\mbox{\boldmath $X$}_{\!\!1n}^{*} - \mbox{\boldmath $X$}_{\!\!2n}^{*} ] \;
  [ {\cal D}^{s}_{\!1m} + i \, {\cal D}^s_{\!6m} ] \; ,
\\
{\cal N}^{\scriptscriptstyle R}_{\!\scriptscriptstyle \tilde{W}\!nm} 
&=&  \frac{1}{{2}} \,       \mbox{\boldmath $X$}_{\!\!3n}^{*} \;
  [ {\cal D}^{s}_{\!1m} + i \, {\cal D}^s_{\!6m} ]\;
- \frac{1}{{2}} \,       
  \mbox{\boldmath $X$}_{\!\!(4+\za)n}^{*}  \;
    [ {\cal D}^{s}_{\!(2+\za)m} - i \, {\cal D}^s_{\!(7+\za)m} ]\; ,
\\
{\cal N}^{\scriptscriptstyle R}_{\!\scriptscriptstyle \tilde{B}\!nm} 
&=& - \frac{1}{{2}} \,       \tan\!\theta_{\!\scriptscriptstyle W} 
\mbox{\boldmath $X$}_{\!\!3n}^{*}  \;
  [ {\cal D}^{s}_{\!1m} + i \, {\cal D}^s_{\!6m} ]\;
+ \frac{1}{{2}} \,        \tan\!\theta_{\!\scriptscriptstyle W} 
\mbox{\boldmath $X$}_{\!\!(4+\za)n}^{*}  \;
  [ {\cal D}^{s}_{\!(2+\za)m} - i \, {\cal D}^s_{\!(7+\za)m} ]\; .
\end{eqnarray}
The list of extra ${\cal N}^{\scriptscriptstyle R^*}$ terms each combines
with  the ${\cal N}^{\scriptscriptstyle R^*}_{\!\scriptscriptstyle inm}$ 
to give a neutral loop contribution to $\Pi_{\xi}$.

\subsection{Charged Loop Contributions}
The effective interaction for the external ${l}_i^{\scriptscriptstyle 0}$ 
with (colorless) charged fermions and scalars inside the loop is given by
\begin{equation}
\label{intC}
{\cal L} = {g_{\scriptscriptstyle 2}}  \,
\overline{\Psi}({l}_i^{\scriptscriptstyle 0})
 \left[ {\cal C}^{\scriptscriptstyle L}_{inm}\, 
{1 - \gamma_{\scriptscriptstyle 5} \over 2} + 
{\cal C}^{\scriptscriptstyle R}_{inm} \, 
{1 + \gamma_{\scriptscriptstyle 5} \over 2} \right ] \, 
 \Psi({\chi}^{\!\!\mbox{ -}}_n) \; \phi_{m}^{\!\scriptscriptstyle +} \;
+ \mbox{h.c.} \;,
\end{equation}
where 
\beqa
{\cal C}^{\scriptscriptstyle R}_{inm} 
&=&
\frac{y_{\!\scriptscriptstyle e_i}}{g_{\scriptscriptstyle 2}} \,
\mbox{\boldmath $V$}_{\!\!(i+2)n} \, D^{l^*}_{\!2m} 
- \frac{\lambda_{ikh}^{\!*}}{g_{\scriptscriptstyle 2}} 
\mbox{\boldmath $V$}_{\!\!(h+2)n} \, D^{l^*}_{\!(k+2)m}  \; ,
\nonumber \\
{\cal C}^{\scriptscriptstyle L}_{inm} 
&=&  
- \mbox{\boldmath $U$}_{\!1n} \, {\cal D}^{l^*}_{\!(i+2)m}
+ \frac{y_{\!\scriptscriptstyle e_i}}{g_{\scriptscriptstyle 2}} \,
  \mbox{\boldmath $U$}_{\!2n} \, {\cal D}^{l^*}_{\!(i+5)m}
- {\lambda_{ihk} \over g_{\scriptscriptstyle 2} } \, 
\mbox{\boldmath $U$}_{\!(h+2)n} \, {\cal D}^{l^*}_{\!(k+5)m} \; .
\label{Cnm}
\eeqa
Here, $\mbox{\boldmath $V$}^\dag \, {\cal M}_{\scriptscriptstyle C} \, \mbox{\boldmath $U$}
= \rm{diag}\{{M}_{\!\scriptscriptstyle \chi^{\mbox{-}}_{n}}\}$ where
${\cal M}_{\scriptscriptstyle C}$ is the $5\times 5$ charged fermion mass matrix.
Matrix ${\cal D}^{l}$ diagonalizes the mass-squared matrix of eight scalars of
unit negative charge (see Ref.\cite{as8} for details). The latter includes again the 
unphysical Goldstone mode to be dropped from the sum over $m$. 

Charged fermion loop contribution to direct 1-loop 
neutrino mass could then be easily obtained as
\beq \label{pinu}
\Pi_{\nu_{ij}}^{\scriptscriptstyle C} = -
{\alpha_{\mbox{\tiny em}} \over 8 \pi \, 
\sin\!^2\theta_{\!\scriptscriptstyle W}} \;
 {\cal C}_{\!\scriptscriptstyle inm}^{\scriptscriptstyle R^*} \,
 {\cal C}_{\!\scriptscriptstyle jnm}^{\scriptscriptstyle L} \;
{M}_{\!\scriptscriptstyle \chi^{\mbox{-}}_{n}} \, 
{\cal B}_{0}\!\!\left(p^2,{M}_{\!\scriptscriptstyle \chi^{\mbox{-}}_{n}}^2,
 {M_{\!\scriptscriptstyle \tilde{\ell}_{m}}^2} \right)
\quad\quad (\, i \,\longleftrightarrow\, j \,) \; .
\eeq
Unlike the case for the neutral loop result, the 
$\Pi_{\nu_{ij}}^{\scriptscriptstyle C}$ matrix written
through the ${\cal C}_{\!\scriptscriptstyle inm}^{\scriptscriptstyle R^*} \,
 {\cal C}_{\!\scriptscriptstyle jnm}^{\scriptscriptstyle L}$ coupled-vertices
is not symmetric with respect to $i$ and $j$. Hence, an explicit 
symmetrization has to be performed, as indicated above. 
The symmetrization also takes care of the asymmetry with respect 
to $L$ and $R$, automatically. 
Similarly, for the $\Sigma_\nu$ part, we have
\beq
\Sigma_{\nu_{ij}}^{\scriptscriptstyle C} = 
{\alpha_{\mbox{\tiny em}} \over 8 \pi \, 
\sin\!^2\theta_{\!\scriptscriptstyle W}} \;
\left\{ \,  {\cal C}_{\!\scriptscriptstyle inm}^{\scriptscriptstyle L} \,
 {\cal C}_{\!\scriptscriptstyle jnm}^{\scriptscriptstyle L^*} \;
\, + \, {\cal C}_{\!\scriptscriptstyle inm}^{\scriptscriptstyle R^*} \,
 {\cal C}_{\!\scriptscriptstyle jnm}^{\scriptscriptstyle R} \; \right \} \;
{\cal B}_{1}\!\!\left(p^2,{M}_{\!\scriptscriptstyle \chi^{\mbox{-}}_{n}}^2,
 {M_{\!\scriptscriptstyle \tilde{\ell}_{m}}^2} \right)
\quad\quad (\, i \,\longleftrightarrow\, j \,) \; . 
\eeq

To go on to discussions of the indirect 1-loop contributions, we
need  the corresponding expressions of the
${\cal C}^{\scriptscriptstyle L,R}_{\!\scriptscriptstyle inm}$ 
for the other four neutral fermions. These are given as follows, 
with obvious notations, 
\beqa
{\cal C}^{\scriptscriptstyle R}_{\!\scriptscriptstyle 0nm} 
&=&
- \frac{y_{\!\scriptscriptstyle e_k}}{g_{\scriptscriptstyle 2}} \,
\mbox{\boldmath $V$}_{\!\!(k+2)n} \, D^{l^*}_{\!(k+2)m} \; ,
\nonumber \\
{\cal C}^{\scriptscriptstyle L}_{\!\scriptscriptstyle 0nm} 
&=&  
- \mbox{\boldmath $U$}_{\!1n} \, {\cal D}^{l^*}_{\!2m}
- \frac{y_{\!\scriptscriptstyle e_k}}{g_{\scriptscriptstyle 2}} \,
  \mbox{\boldmath $U$}_{\!(k+2)n} \, {\cal D}^{l^*}_{\!(k+5)m}\; ;
\\
{\cal C}^{\scriptscriptstyle R}_{\!\scriptscriptstyle \tilde{h}\!nm} 
&=&
- \mbox{\boldmath $V$}_{\!\!1n} \, {\cal D}^{l^*}_{\!1m} \; ,
\nonumber \\
{\cal C}^{\scriptscriptstyle L}_{\!\scriptscriptstyle \tilde{h}\!nm} 
&=&  
0 \; ;
\\
{\cal C}^{\scriptscriptstyle R}_{\!\scriptscriptstyle \tilde{W}\!nm} 
&=&
 \frac{-1}{\sqrt{2}} \, 
\mbox{\boldmath $V$}_{\!\!2n} \, {\cal D}^{l^*}_{\!1m} \; ,
\nonumber \\
{\cal C}^{\scriptscriptstyle L}_{\!\scriptscriptstyle \tilde{W}\!nm} 
&=&  
\frac{1}{\sqrt{2}} \, \left[
\mbox{\boldmath $U$}_{\!2n} \, {\cal D}^{l^*}_{\!2m}
+ \mbox{\boldmath $U$}_{\!(k+2)n} \, 
{\cal D}^{l^*}_{\!(k+2)m} 
\right] \; ;
\label{CWnm}\\
{\cal C}^{\scriptscriptstyle R}_{\!\scriptscriptstyle \tilde{B}\!nm} 
&=&
\frac{-\tan\!\theta_{\!\scriptscriptstyle W}}{\sqrt{2}} \, \left[
\mbox{\boldmath $V$}_{\!\!2n} \, {\cal D}^{l^*}_{\!1m}
+ 2\, \mbox{\boldmath $V$}_{\!\!(k+2)n} \, 
{\cal D}^{l^*}_{\!(k+5)m} 
\right]\; ,
\nonumber \\
{\cal C}^{\scriptscriptstyle L}_{\!\scriptscriptstyle \tilde{B}\!nm} 
&=&  
\frac{\tan\!\theta_{\!\scriptscriptstyle W}}{\sqrt{2}} \, \left[
\mbox{\boldmath $U$}_{\!2n} \, {\cal D}^{l^*}_{\!2m}
+ \mbox{\boldmath $U$}_{\!(k+2)n} \, 
{\cal D}^{l^*}_{\!(k+2)m} 
\right] \; .
\label{CBnm}
\end{eqnarray}
Combining a ${\cal C}^{\scriptscriptstyle R^*}$ with a 
${\cal C}^{\scriptscriptstyle L}$ gives half of the charged fermion loop 
contribution, to the corresponding mass term; the other half is given by 
flipping $L$ and $R$. For instance, the 
${l}_i^{\scriptscriptstyle 0}$-$\tilde{B}$ mass term, or
$\Pi_{\xi_{i1}}$, is given by substituting  
${\cal C}^{\scriptscriptstyle L}_{\!\scriptscriptstyle \tilde{B}\!nm}$ for
${\cal C}^{\scriptscriptstyle L}_{\!\scriptscriptstyle jnm}$ in Eq.(\ref{pinu}), {\it i.e.,}
by a ${\cal C}_{\!\scriptscriptstyle inm}^{\scriptscriptstyle R^*} \,
{\cal C}^{\scriptscriptstyle L}_{\!\scriptscriptstyle \tilde{B}\!nm}$ 
combination, as well as the combination
${\cal C}_{\!\scriptscriptstyle \tilde{B}\!nm}^{\scriptscriptstyle R^*} \,
{\cal C}^{\scriptscriptstyle L}_{\!\scriptscriptstyle inm}$. 

There is also another type of contributions, namely the quark-squark loops.
The direct 1-loop part of such contributions is among the most well discussed. 
We summarize them here, under our notation, for completeness. We have
\beq \label{dpinu}
\Pi_{\nu_{ij}}^{\scriptscriptstyle D} = -
{\alpha_{\mbox{\tiny em}} {N}_{c} \over 8 \pi \, 
\sin\!^2\theta_{\!\scriptscriptstyle W}} \;
 {\cal C}_{\!\scriptscriptstyle inm}^{\prime\scriptscriptstyle R^*} \,
 {\cal C}_{\!\scriptscriptstyle jnm}^{\prime\scriptscriptstyle L} \;
{m}_{\!\scriptscriptstyle d_{n}} \, 
{\cal B}_{0}\!\!\left(p^2,{m}_{\!\scriptscriptstyle d_{n}}^2,
 {M_{\!\scriptscriptstyle \tilde{d}_{m}}^2} \right)
\quad\quad (\, i \,\longleftrightarrow\, j \,) \; 
\eeq
where 
\beqa
{\cal C}^{\prime\scriptscriptstyle R}_{inm} 
&=&
- \frac{\lambda_{\scriptscriptstyle ikn}^{\prime \, \!*}}
       {g_{\scriptscriptstyle 2}} \, {\cal D}^{d^*}_{\!km}  \; ,
\nonumber \\
{\cal C}^{\prime\scriptscriptstyle L}_{inm} 
&=&  
- {\lambda_{\scriptscriptstyle ink}^{\prime} \over g_{\scriptscriptstyle 2} } 
        \, {\cal D}^{d^*}_{\!(k+3)m} \;, 
\label{Cpnm}
\eeqa
and ${\cal D}^d$ diagonalizes the $6\times 6$ squark mass-squared matrix 
${\cal M}_{\!\scriptscriptstyle D}^2$. The structure is to be compared
directly with those from the $\lambda$-couplings above. For $\Sigma_\nu$,
we have
\beq
\Sigma_{\nu_{ij}}^{\scriptscriptstyle D} = 
{\alpha_{\mbox{\tiny em}} \over 8 \pi \, 
\sin\!^2\theta_{\!\scriptscriptstyle W}} \;
\left\{ \,  {\cal C}_{\!\scriptscriptstyle inm}^{\prime\scriptscriptstyle L} \,
 {\cal C}_{\!\scriptscriptstyle jnm}^{\prime\scriptscriptstyle L^*} \;
\, + \, {\cal C}_{\!\scriptscriptstyle inm}^{\prime\scriptscriptstyle R^*} \,
 {\cal C}_{\!\scriptscriptstyle jnm}^{\prime\scriptscriptstyle R} \; \right \} \;
{\cal B}_{1}\!\!\left(p^2,{M}_{\!\scriptscriptstyle d_{n}}^2,
 M_{\!\scriptscriptstyle \tilde{d}_{m}}^2 \right)
\quad\quad (\, i \,\longleftrightarrow\, j \,) \; .
\eeq
For the indirect 1-loop part, we need
\beqa
{\cal C}^{\prime\scriptscriptstyle R}_{\!\scriptscriptstyle 0nm} 
&=&
- \frac{y_{\!\scriptscriptstyle d_n}}{g_{\scriptscriptstyle 2}} \,
 {\cal D}^{d^*}_{\!km} \; ,
\nonumber \\
{\cal C}^{\prime\scriptscriptstyle L}_{\!\scriptscriptstyle 0nm} 
&=&  
- \frac{y_{\!\scriptscriptstyle d_n}}{g_{\scriptscriptstyle 2}} \,
{\cal D}^{d^*}_{\!(k+3)m}\; ;
\\
{\cal C}^{\prime\scriptscriptstyle R}_{\!\scriptscriptstyle \tilde{W}\!nm} 
&=&
0 \; ,
\nonumber \\
{\cal C}^{\prime\scriptscriptstyle L}_{\!\scriptscriptstyle \tilde{W}\!nm} 
&=& 
\frac{1}{\sqrt{2}} \, {\cal D}^{d^*}_{\!km} \; ;
\\
{\cal C}^{\prime\scriptscriptstyle R}_{\!\scriptscriptstyle \tilde{B}\!nm} 
&=&
-\frac{\sqrt{2}}{3} \, \tan\!\theta_{\!\scriptscriptstyle W} \,
   {\cal D}^{d^*}_{\!(k+3)m} \; ,
\nonumber \\
{\cal C}^{\prime\scriptscriptstyle L}_{\!\scriptscriptstyle \tilde{B}\!nm} 
&=&  
-\frac{\sqrt{2}}{6} \, \tan\!\theta_{\!\scriptscriptstyle W} \,
 {\cal D}^{d^*}_{\!km} \; .
\end{eqnarray}
 We get the indirect 1-loop
contributions by combining ${\cal C}^{\prime\scriptscriptstyle R(L)^*}$
with  ${\cal C}^{\prime\scriptscriptstyle L(R)}_{\scriptscriptstyle inm}$,
in the same way as we do in the above case of (colorless) charged fermion loop.

\section{More Detailed Analytical Results}
In this section, we give more explicit details of the neutrino mass terms obtained
by applying the formulae in the previous section. We list the result from different
combinations of interaction vertices and go on to illustrate the content of these 
exact mass eigenstate results by extracting the dominating piece(s) in the mass
eigenstate double sum. There, we give the ``approximate" analytical results
through the use of perturbative diagonalization expressions\cite{as8} for the 
elements of the various mixing matrices. Such perturbative diagonalizations
have been illustrated to be very good approximations, which also serve to
illustrate well the role of the various lepton flavor violating (LFV) couplings
involved (see Refs.\cite{as46,as7} for other illustrations).

Note that we focus our discussions below only on the parts of the results that
are particularly interesting to our analytical study. For instance, in the 
$\Pi_{\nu_{ij}}^{\scriptscriptstyle N} = -
{\alpha_{\mbox{\tiny em}} \over 8 \pi \, 
\sin\!^2\theta_{\!\scriptscriptstyle W}} \;
 {\cal N}_{\!\scriptscriptstyle inm}^{\scriptscriptstyle R^*} \,
 {\cal N}_{\!\scriptscriptstyle jnm}^{\scriptscriptstyle R^*} \;
{M}_{\!\scriptscriptstyle \chi^0_{n}} \, 
{\cal B}_{0}\!\!\left(p^2,{M}_{\!\scriptscriptstyle \chi^0_{n}}^2,
 {M_{\!\scriptscriptstyle S_{m}}^2} \right)$ term, we focus on the 
$ {\cal N}_{\!\scriptscriptstyle inm}^{\scriptscriptstyle R^*} \,
 {\cal N}_{\!\scriptscriptstyle jnm}^{\scriptscriptstyle R^*} \;
{M}_{\!\scriptscriptstyle \chi^0_{n}} $
part. That is, we will drop the common pre-factor 
${\alpha_{\mbox{\tiny em}} \over 8 \pi \, 
\sin\!^2\theta_{\!\scriptscriptstyle W}}$ and the loop integral ${\cal B}_{0}$
from all the neutrino mass term results given below.
The following discussions do not include the $\Sigma_\nu$ part. The results
of the latter are left all to an appendix at the end. They are included here 
mainly for completeness. It does not look like there is any important
off-diagonal contribution, while diagonal contributions, as discussed above,
only represent a universal correction to the tree-level result.

\subsection{Results for \boldmath $\Pi_{\nu_{ij}}^{\scriptscriptstyle N}$ 
and $(\Pi_\xi \,  {\cal M}_n^{\mbox{-}1} \,  \xi^{\!\scriptscriptstyle  T})
^{\scriptscriptstyle N}_{ij}$ } 
\begin{itemize}
\item
The result here may be written in the form of a single term as
\beqa
&&
\frac{1}{{4}} \,       
[ \tan\!\theta_{\!\scriptscriptstyle W} 
\mbox{\boldmath $X$}_{\!\!1n} - \mbox{\boldmath $X$}_{\!\!2n} ]^2 \,
{M}_{\!\scriptscriptstyle \chi^0_{n}} \;
  [ {\cal D}^{s}_{\!(i+2)m} - i \, {\cal D}^s_{\!(i+7)m} ] \;
[ {\cal D}^{s}_{\!(j+2)m} - i \, {\cal D}^s_{\!(j+7)m} ]
\nonumber \\ 
&\simeq&
\frac{B_i B_j\, \tan\!^2\!\zb }{M_{\!s}^3} \; 
\left[ \frac{1}{{4}} \,  (\tan\!\theta_{\!\scriptscriptstyle W} 
\mbox{\boldmath $X$}_{\!\!1n} - \mbox{\boldmath $X$}_{\!\!2n} )^2 \,
\frac{{M}_{\!\scriptscriptstyle \chi^0_{n}} }{M_{\!s}} \right] \,
 \nonumber \\ 
&& \mbox{\hspace*{4in} (n=1-4 dominate)} \;.
\eeqa
The scalar sum is dominated by $m=1,2$ and $7$ contributions. We illustrate here
only the dependence on the $B_i$ parameters and $\tan\!\zb$, with ${M_{\!s}}$ 
denoting a generic mass parameter at the slepton scale.  
$\!\!\!$\footnote{  
Note that we have  $[ {\cal D}^{s}_{\!(i+2)1} - i \, {\cal D}^s_{\!(i+7)1} ] \simeq
\frac{-\rm{Re}(B_i) }{M_{\!s}^2} -i \,\frac{\rm{Im}(B_i) }{M_{\!s}^2}
\, [\tan\!\zb \, \sin\!\za - \cos\!\za] = - \frac{B_i }{M_{\!s}^2} \, [\tan\!\zb \, \sin\!\za - \cos\!\za]$ 
from our perturbative formulae on the ${\cal D}^{s}$ elements.  One may also 
check the other pieces. Take the $m=2$ piece, for example, we have then 
$[ {\cal D}^{s}_{\!(i+2)1} - i \, {\cal D}^s_{\!(i+7)1} ] \simeq
- \frac{B_i }{M_{\!s}^2} \, [\tan\!\zb \, \cos\!\za + \sin\!\za]$; for  $m=7$,  
$[ {\cal D}^{s}_{\!(i+2)7} - i \, {\cal D}^s_{\!(i+7)7} ] \simeq
\frac{-\rm{Im}(B_i) +i \,\rm{Re}(B_i)}{cos\zb \, M_{\!s}^2}
=i \, \frac{B_i }{M_{\!s}^2} \, \frac{1}{cos\zb}$. 
The extra factor of $i$ guarantees a cancelation with the $m=1$ and $2$ terms 
if the ($m=7$) ``pseudoscalar" is mass degenerate with the latter ``scalars", as 
$ [\tan\!\zb \, \sin\!\za - \cos\!\za]^2 +[\tan\!\zb \, \cos\!\za + \sin\!\za]^2 =
\frac{1}{cos^2\zb}$. Hence, to illustrate the generic result, we write the dominating
$[ {\cal D}^{s}_{\!(i+2)m} - i \, {\cal D}^s_{\!(i+7)m} ] $ result as 
$\frac{-B_i \, \tan\!\zb}{M_{\!s}^2}$. This is used throughout the section.
}
Note that we write the final result in the form such that the square-bracket 
[~] contains a factor of order 1 (a pattern we stick to below), so that the reader
can have an idea on the major parameters (those before the square-bracket) 
affecting the scale of the neutrino mass.  The resultant proportionality of the 
mass term here to the product $B_i B_j$ has been addressed and interpreted as
the necessity for a Majorana-like scalar mass insertion to complete the diagram,
in terms of complex scalars. When one follows such an interpretation to consider
the scalar inside the loop as complex field with mass insertions
put in on the line explicitly  (as shown in Fig.~6 of Ref.\cite{as5} for example),
a proportionality on $B_i B_j$ would likewise be resulted. The different $m$ pieces
in the scalar sum, however, cannot be put together at this level. Each piece involves
actually a different value for $M_{\!s}$ and a different loop integral from
a physical scalar of different mass running in the loop. 
In fact, if one naively takes a sum over $m$ without considering the loop
integrals, a zero result would be obtained for any 
${\cal D}^{s}_{\!am}{\cal D}^{s}_{\!bm}$ with $a\ne b$. The lack of
degeneracy among the scalar mass eigenstates is what makes a nonzero result
possible. This is a common feature for the type of diagrams (see also Ref.\cite{as7}).
Interestingly enough, for the present case under discussion, a pairwise degeneracy
among the ``scalar" and ``pseudoscalar" parts of a complex scalar is enough
to guarantee a null result. This is equivalent to the statement that the neutrino
mass contribution is proportional to a Majorana-like mass term. It is illustrated
here in our expressions as a consequence of the cancelation
between ${\cal D}^{s}_{\!(i+2)m}{\cal D}^{s}_{\!(j+2)m}$ and 
$i^2\,{\cal D}^{s}_{\!(i+7)m}{\cal D}^{s}_{\!(j+7)m}$ as well as between 
${\cal D}^{s}_{\!(i+2)m}{\cal D}^{s}_{\!(j+7)m}$ and
${\cal D}^{s}_{\!(i+7)m}{\cal D}^{s}_{\!(j+2)m}$ for each single $m$ value,
from our perturbative expressions for the mixing matrix elements.
\end{itemize}

Next, we come to the 
$(\Pi_\xi \,  {\cal M}_n^{\mbox{-}1} \,  \xi^{\!\scriptscriptstyle  T})^{\scriptscriptstyle N}_{ij}$
part. The dominating results from all the individual terms of the form
have a common proportionality to the combination of LFV parameters
\[
{B_i \, \mu_j \; (\tan\!\zb)} \;.
\]
Again, the contribution mainly involves diagrams with a (physical) neutralino,
together with a neutral scalar, running in the loop. As noted above, the lack
of mass degeneracy among the scalars is essential for a nontrivial result. 
Note that upon the necessary symmetrization not explicitly shown, we will
have also the ${B_j \, \mu_i \; (\tan\!\zb)}$ parameter combination coming in.

All the different terms in this class have very similar structure.
We discuss here only an illustrative term, and leave the rest to Appendix A
below. Let us take a look at the term
 $\Pi_{\nu_i \!0}^{\scriptscriptstyle N} \, ({\cal M}_n^{\mbox{-}1} \,  \xi^{\!\scriptscriptstyle  T})_{4j}$.
It is given as
\footnotesize \beqa 
& & - \mu_j \; \frac{\mu_{\scriptscriptstyle 0}  
M_{\!\scriptscriptstyle 1}   M_{\!\scriptscriptstyle 2} 
- M_{\!\scriptscriptstyle Z}^2 \, \sin\!\beta \cos\!\beta \,
 (M_{\!\scriptscriptstyle 1} \cos\!^2 \theta_{\!\scriptscriptstyle W}
 + M_{\!\scriptscriptstyle 2} \sin\!^2\theta_{\!\scriptscriptstyle W})}
{\det({\cal M}_n)} \; 
\frac{1}{{4}} \,       [ \tan\!\theta_{\!\scriptscriptstyle W} 
\mbox{\boldmath $X$}_{\!\!1n} - \mbox{\boldmath $X$}_{\!\!2n} ]^2 \,
{M}_{\!\scriptscriptstyle \chi^0_{n}} \;
 \nonumber \\
&&  \mbox{\hspace*{3in} }\cdot
 [ {\cal D}^{s}_{\!(i+2)m} - i \, {\cal D}^s_{\!(i+7)m} ] \;
[ {\cal D}^{s}_{\!2m} - i \, {\cal D}^s_{\!7m} ]
\nonumber \\[.2in] 
&\simeq& 
\frac{B_i \, \mu_j \, \tan\!\zb}{M_{\!s}^2} \,
\left[ \frac{\mu_{\scriptscriptstyle 0} 
M_{\!\scriptscriptstyle 1}  M_{\!\scriptscriptstyle 2} 
- M_{\!\scriptscriptstyle Z}^2 \, \sin\!\beta \cos\!\beta \,
 (M_{\!\scriptscriptstyle 1} \cos\!^2 \theta_{\!\scriptscriptstyle W}
 + M_{\!\scriptscriptstyle 2} \sin\!^2\theta_{\!\scriptscriptstyle W})}{\det({\cal M}_n)} \; 
\frac{1}{4} \,   (\tan\!\theta_{\!\scriptscriptstyle W} 
\mbox{\boldmath $X$}_{\!\!1n} - \mbox{\boldmath $X$}_{\!\!2n} )^2 \,
{M}_{\!\scriptscriptstyle \chi^0_{n}} \right]\;.
\label{N4j}
\eeqa \normalsize
Note that from the general flavor structure of the model, one expect 
$\Pi_{\nu_i \!0}^{\scriptscriptstyle N}$ to have an expression similar to 
$\Pi_{\nu_{ij}}^{\scriptscriptstyle N}$ above with index $j$ replaced by a $0$, 
{\it i.e.}, $\Pi_{\nu_i \!0}^{\scriptscriptstyle N} \simeq \frac{B_i B_0}{M_{\!s}^4} \;
 \tan\!^2\!\zb \; \frac{1}{{4}} \,  ( \tan\!\theta_{\!\scriptscriptstyle W} 
\mbox{\boldmath $X$}_{\!\!1n} - \mbox{\boldmath $X$}_{\!\!2n} )^2 \,
{M}_{\!\scriptscriptstyle \chi^0_{n}} $. 
Observing that $\frac{B_0}{\cos\!\zb}$ is a parameter of the same order as 
the generic mass scale parameter $M_{\!s}^2 \, \tan\!\zb$ 
we do see an agreement here. 

\subsection{Results for the Charged and Color Loops} 
Similar to the neutral loop case above, each term in the charged loop contributions 
to the $\Pi$'s has a scalar part involving 
$D^{l}_{\!am} \, {\cal D}^{l^*}_{\!bm}$ which would give a null result
for $a\ne b$ if summed over $m$ naively. The different loop integrals from the 
lack of scalar degeneracy is what guarantees nontrivial results. The fermionic 
part is more interesting. For illustrative purpose, we take an expression of 
the form    $\mbox{\boldmath $V$}_{\!\!(i+2)n}^* \, 
{M}_{\!\scriptscriptstyle \chi^{\mbox{-}}_{n}} \, \mbox{\boldmath $U$}_{\!1n}$.
Here, $n=1$ and $2$ give the chargino state contributions, with a large
${M}_{\!\scriptscriptstyle \chi^{\mbox{-}}_{n}}$ mass but a more suppressed
$\mbox{\boldmath $V$}_{\!\!(i+2)n}^*$ mixing. The results are given by 
$\frac{ R_{{\!\scriptscriptstyle R}_{21}}^*}{M_{c{\scriptscriptstyle 1}}} 
\,{m_i\,\mu_i} \, \mbox{\boldmath $U$}_{\!1n}$ and
$\frac{ R_{{\!\scriptscriptstyle R}_{22}}^*}{M_{c{\scriptscriptstyle 2}}} 
 \,{m_i\,\mu_i}\, \mbox{\boldmath $U$}_{\!2n}$,
respectively. On the other hand, the $n=i+2$ term involves a small
fermion mass ${M}_{\!\scriptscriptstyle \chi^{\mbox{-}}_{i+2}} = m_i$
but a less suppressed mixing of $\mbox{\boldmath $U$}_{\!1(i+2)} \simeq 
\frac{ \sqrt{2}\, M_{\!\scriptscriptstyle W} \cos\!\beta}{M_{\!\scriptscriptstyle 0}^2} \,\mu_i$.
Dropping all the factors of order 1, we have all three terms giving contribution
of roughly the same order of magnitude, all proportional to $\frac{m_i \, \mu_i }{M_{s}}$,
where ${M_{s}}$ again denotes a SUSY scale mass parameter here corresponds, more
exactly, to a chargino mass. This kind of feature is quite common in the charged loop
results below. We illustrate results by dropping all the order 1 parameters and using
the generic mass parameter ${M_{s}}$ representing chargino as well as slepton
mass scale. 

There are six terms to the  $\Pi_{\nu_{ij}}^{\scriptscriptstyle C}$ result. We 
mostly just list them, while drawing attention to particularly interesting 
features. Note that the necessary symmetrization is not shown explicitly.
\begin{itemize}
\item ~ \vspace*{-.3in}
\beq
- \frac{y_{\!\scriptscriptstyle e_i}}{g_{\scriptscriptstyle 2}} \,
\mbox{\boldmath $V$}_{\!\!(i+2)n}^* \, 
{M}_{\!\scriptscriptstyle \chi^{\mbox{-}}_{n}} \, \mbox{\boldmath $U$}_{\!1n} \;
D^{l}_{\!2m} \, {\cal D}^{l^*}_{\!(j+2)m}
\simeq
- \frac{y_{\!\scriptscriptstyle e_i}}{g_{\scriptscriptstyle 2}} \,
\frac{m_i \, \mu_i \, B_j \, \tan\!\zb}{M^3_{s}}
\eeq
The scalar part result here is mainly from $D^{l}_{\!2(2+j)} \simeq 
\frac{B_j \, \tan\!\zb}{M^2_{s}}$.
\item ~ \vspace*{-.3in}
\beq
\frac{y_{\!\scriptscriptstyle e_i}}{g_{\scriptscriptstyle 2}} \,
\frac{y_{\!\scriptscriptstyle e_j}}{g_{\scriptscriptstyle 2}} \,
\mbox{\boldmath $V$}_{\!\!(i+2)n}^* \,  
 {M}_{\!\scriptscriptstyle \chi^{\mbox{-}}_{n}} \, 
\mbox{\boldmath $U$}_{\!2n} \;
D^{l}_{\!2m} \, {\cal D}^{l^*}_{\!(j+5)m}
\simeq
\frac{y_{\!\scriptscriptstyle e_i}}{g_{\scriptscriptstyle 2}} \,
\frac{y_{\!\scriptscriptstyle e_j}}{g_{\scriptscriptstyle 2}} \,
\frac{m_i \, m_j \, \mu_i  \, \mu_j  \, \tan\!\zb}{M^3_{s}}
\eeq
The scalar part result here is mainly from $D^{l}_{\!2(5+j)} \simeq 
\frac{m_j \, \mu_j  \tan\!\zb}{M^2_{s}}$.
\item ~ \vspace*{-.3in}
\ 
\beqa
 - \frac{y_{\!\scriptscriptstyle e_i}}{g_{\scriptscriptstyle 2}} \,
{\lambda_{jhk} \over g_{\scriptscriptstyle 2} } \, 
\mbox{\boldmath $V$}_{\!\!(i+2)n}^* \, 
{M}_{\!\scriptscriptstyle \chi^{\mbox{-}}_{n}} \, 
\mbox{\boldmath $U$}_{\!(h+2)n} \;
D^{l}_{\!2m} \, {\cal D}^{l^*}_{\!(k+5)m}  
\simeq  
- \frac{y_{\!\scriptscriptstyle e_i}}{g_{\scriptscriptstyle 2}} \,
 {\lambda_{jik} \over g_{\scriptscriptstyle 2} } \,
\frac{m_i \, m_k \, \mu_k  \, \tan\!\zb}{M^2_{s}} 
\eeqa
Here, the result is from $n=i+2$ which is interesting only at $h=i$; hence,
only that is shown in the sum over $h$. It is the SUSY analog of the Zee
diagram, discussed in Refs.\cite{as1,as5}.
For $h\ne i$ parts, the result is much further 
suppressed (by another $\frac{\mu_i  \, \mu_h^* } {M^2_{s}}$ factor).  
The scalar part result is the same as the previous case.
\item ~ \vspace*{-.3in}
\ 
\beq \label{mu-lam}
\frac{\lambda_{ikh}}{g_{\scriptscriptstyle 2}} \,
\mbox{\boldmath $V$}_{\!\!(h+2)n}^* \,  
{M}_{\!\scriptscriptstyle \chi^{\mbox{-}}_{n}} \, \mbox{\boldmath $U$}_{\!1n} \;
 D^{l}_{\!(k+2)m} \, {\cal D}^{l^*}_{\!(j+2)m}
\simeq
\frac{\lambda_{ijh}}{g_{\scriptscriptstyle 2}} \,
\frac{m_h \, \mu_h}{M_{s}}
\qquad \quad \mbox{(symmetrization !)}
\eeq
We note here that the result is actually very sensitive to the $i\leftrightarrow j$ 
symmetrization. The dominant result in the expression above  is from the case with the
$(j+2)$th charged scalar running in the loop. This is approximately the 
$\tilde{l}^{\!\!\mbox{ -}}_j$ slepton. The symmetrization and the fact that
$\lambda_{ijh}=-\lambda_{jih}$ suggest a perfect cancelation of the result in 
the limit of degenerate sleptons which correspond roughly to the   
$\tilde{l}^{\!\!\mbox{ -}}_i$ and $\tilde{l}^{\!\!\mbox{ -}}_j$ 
states. This has also been discussed in some detail in Ref.\cite{as5}.
\item ~ \vspace*{-.3in}
\ 
\beq
- \frac{y_{\!\scriptscriptstyle e_j}}{g_{\scriptscriptstyle 2}} \,
\frac{\lambda_{ikh}}{g_{\scriptscriptstyle 2}} \,
\mbox{\boldmath $V$}_{\!\!(h+2)n}^* \,  
{M}_{\!\scriptscriptstyle \chi^{\mbox{-}}_{n}} \,  \mbox{\boldmath $U$}_{\!2n} \;
D^{l}_{\!(k+2)m}  \,  {\cal D}^{l^*}_{\!(j+5)m}
\simeq
- \frac{y_{\!\scriptscriptstyle e_j}}{g_{\scriptscriptstyle 2}} \,
\frac{\lambda_{ikh}}{g_{\scriptscriptstyle 2}} \,
\frac{m_h \, \mu_h\, (\widetilde{\cal M}_{\!\scriptscriptstyle RL}^{2})^*_{jk} }{M_{s}^3} \; 
\eeq
where
\beq
(\widetilde{\cal M}_{\!\scriptscriptstyle RL}^{2})_{jk} 
=  [A_e^* - \mu_{\scriptscriptstyle 0} \, \tan\!\beta ] \, m_j \, \delta_{kj}
+ \frac{\sqrt{2}\, M_{\!\scriptscriptstyle W} \cos\!\beta}
{g_{\scriptscriptstyle 2} } \,   \delta\! A^{\!{\scriptscriptstyle E}^*}_{kj}
 - \frac{\sqrt{2}\, M_{\!\scriptscriptstyle W} \sin\!\beta}
{g_{\scriptscriptstyle 2} } \,(\, \mu_i\lambda_{ikj}^*\, )  \; .
\eeq
gives the complete  $LR$ mixing of  $\tilde{l}^{\scriptscriptstyle +}_j$
and $\tilde{l}^{\!\!\mbox{ -}}_k$ states. The last part of the latter is a 
contribution beyond the well known MSSM parts.
\item ~ \vspace*{-.3in}
\ 
\beqa
\frac{\lambda_{ikh}}{g_{\scriptscriptstyle 2}} \,
{\lambda_{jqp} \over g_{\scriptscriptstyle 2} } \, 
 \mbox{\boldmath $V$}_{\!\!(h+2)n}^* \, 
{M}_{\!\scriptscriptstyle \chi^{\mbox{-}}_{n}} \, 
\mbox{\boldmath $U$}_{\!(q+2)n} \; 
D^{l}_{\!(k+2)m}  \, {\cal D}^{l^*}_{\!(p+5)m}
\simeq  \frac{\lambda_{ikh}}{g_{\scriptscriptstyle 2}} \,
{\lambda_{jhp} \over g_{\scriptscriptstyle 2} } \,
\frac{m_h \, (\widetilde{\cal M}_{\!\scriptscriptstyle RL}^{2})^*_{pk}}{M_{s}^2} \; 
\eeqa
This is the most well known part of $\Pi_{\nu_{ij}}^{\scriptscriptstyle C}$ result 
discussed extensively in the literature. Note again the extra (last) term in the $LR$ 
mixing $(\widetilde{\cal M}_{\!\scriptscriptstyle RL}^{2})^*_{pk}$. 
Its contribution to neutrino masses in the case of $p\ne k$ 
may be particularly interesting.\\
\end{itemize}

For the  $(\Pi_\xi \,  {\cal M}_n^{\mbox{-}1} \,  \xi^{\!\scriptscriptstyle  T})
^{\scriptscriptstyle C}_{ij}$ part, we present the long list of terms in Appendix~B.
In the neutral loop counterpart above, we see that
the class of indirect 1-loop result all involve the combination
$B_i\,\mu_j\,\tan\!\zb$. Here for the charged loop results, we see the same
parameter combination does give some important terms, but without the $\tan\!\zb$ 
factor. These are labeled as
$\Pi_{\nu_i \!0}^{\scriptscriptstyle C} \, ({\cal M}_n^{\mbox{-}1} \,
 \xi^{\!\scriptscriptstyle T})_{3j} $ --- part 1 and
$\Pi_{\nu_i \!0}^{\scriptscriptstyle C} \, ({\cal M}_n^{\mbox{-}1} \,
 \xi^{\!\scriptscriptstyle T})_{1j} $ --- part 5 [with corresponding 
$\Pi_{\nu_i \!0}^{\scriptscriptstyle C} \, ({\cal M}_n^{\mbox{-}1} \,
 \xi^{\!\scriptscriptstyle T})_{2j} $ part] inside the appendix.  In factor, these
terms could easily dominate over the direct 1-loop terms from
$\Pi_\nu^{\scriptscriptstyle C}$ over. They provide neutrino mass contributions
of order $\frac{B_i\,\mu_j}{M_s^2}$.

Another type of interesting terms are given by those labeled as
$\Pi_{\nu_i \!0}^{\scriptscriptstyle C} \, ({\cal M}_n^{\mbox{-}1} \,
 \xi^{\!\scriptscriptstyle T})_{1j} $ --- part 4 [again with corresponding 
$\Pi_{\nu_i \!0}^{\scriptscriptstyle C} \, ({\cal M}_n^{\mbox{-}1} \,
 \xi^{\!\scriptscriptstyle T})_{2j} $ part] and
$\Pi_{\nu_i \!0}^{\scriptscriptstyle C} \, ({\cal M}_n^{\mbox{-}1} \,
 \xi^{\!\scriptscriptstyle T})_{1j} $ --- part 10  inside the appendix. 
We have, roughly, the results  
$\frac{\lambda_{ikh}}{g_{\scriptscriptstyle 2}} \,{\mu_j \, m_h \over M_s^2}$
or $\frac{\lambda_{ihk}}{g_{\scriptscriptstyle 2}} \,{\mu_j \, m_h \over M_s^2}$.

Most of the other terms are actually not very interesting. They typically involve
further suppression from factors such as ${m_i \over M_s} 
\frac{y_{\!\scriptscriptstyle e_i}}{g_{\scriptscriptstyle 2}}$. However, one 
should note that for the large $\tan\!\zb$ case, the $i=3$ part has an order
1 coupling (essentially the $\tau$ Yukawa) which renders the suppression
not too strong. A careful numerical study will be necessary to check if 
there could be a scenario where such term could play a role. 

The quark-squark loop results are much more simple as a class. In fact, 
parallel structure between the $\lambda^{\prime}_{ijk}$ coupling terms and
the $\lambda_{ijk}$ coupling terms can also be used to write down the results 
directly. In particular, for the indirect 1-loop part, we expected  
$\frac{\lambda_{ikh}^{\prime}}{g_{\scriptscriptstyle 2}} \,{\mu_j \, m_{d_h} \over M_s^2}$
or $\frac{\lambda_{ihk}^{\prime}}{g_{\scriptscriptstyle 2}} \,{\mu_j \, m_{d_h} \over M_s^2}$
to match the similar terms just discussed above. We list the details in Appendix~C.

\section{Concluding Remarks}
We have listed and discussed the detailed results of all the neutrino mass terms
within the GSSM, up to 1-loop order. Our approach gives expression for exact 
results, each to be obtained through a double summation over the fermion and scalar 
mass eigenstates running inside the loop. We further give approximate 
expressions of each of these terms through extracting the dominating pieces
within the double summation and approximating the elements of the mass mixing 
matrices by perturbative diagonalization formulae. The validity of such 
perturbative diagonalizations are well founded on the experimental smallness
of effects involving lepton flavor violation or R-parity violation. However, there
are partial cancellations among pieces within the sum ---- a result of a GIM type
unitary cancellation also pointed out in Refs.\cite{as46,as7}, rendering the
approximate formulae agree only at order of magnitude level with the summed
exact results.  The latter is also cross-checked through numerical calculations,
part of which is given in Appendix E for illustrative purposes. We most probably
have given the results in more details than necessary. However, we emphasize
that our ignorance about the nature of SUSY parameters, R-parity violating
or otherwise, says that imposing much theoretical prejudice on the likely importance
on some contribution over the others may be unwise. The detailed listing here
is intended to provide a reference to later studies on any plausibly interesting
scenario out the model. Numerical studies of the latter will be published
independently.

\acknowledgements
We would like to thank E.J.~Chun for discussions, and Y.-Y.~Keum for making
available some routines for numerical matrix diagonalization. 
Work of O.~K. has been partially supported by the National Science Council
of Taiwan under grants number {NSC} 91-2112-{M}-008-042 and
{NSC} 90-2112-{M}-008-051, and has benefitted from particle physics program
activities and support of the National Center of Theoretical Sciences of Taiwan.

\appendix
\section{Details of \boldmath
$(\Pi_\xi \,  {\cal M}_n^{\mbox{-}1} \,  \xi^{\!\scriptscriptstyle  T})
^{\scriptscriptstyle N}$ Terms}
\begin{itemize}
\item $\Pi_{\nu_i \!0}^{\scriptscriptstyle N} \, ({\cal M}_n^{\mbox{-}1} \,  \xi^{\!\scriptscriptstyle  T})_{4j}$ --- 
\footnotesize \beqa
& & - \mu_j \; \frac{\mu_{\scriptscriptstyle 0}  
M_{\!\scriptscriptstyle 1}   M_{\!\scriptscriptstyle 2} 
- M_{\!\scriptscriptstyle Z}^2 \, \sin\!\beta \cos\!\beta \,
 (M_{\!\scriptscriptstyle 1} \cos\!^2 \theta_{\!\scriptscriptstyle W}
 + M_{\!\scriptscriptstyle 2} \sin\!^2\theta_{\!\scriptscriptstyle W})}
{\det({\cal M}_n)} \; 
\frac{1}{{4}} \,       [ \tan\!\theta_{\!\scriptscriptstyle W} 
\mbox{\boldmath $X$}_{\!\!1n} - \mbox{\boldmath $X$}_{\!\!2n} ]^2 \,
{M}_{\!\scriptscriptstyle \chi^0_{n}} \;
 \nonumber \\
&&  \mbox{\hspace*{3in} }\cdot
 [ {\cal D}^{s}_{\!(i+2)m} - i \, {\cal D}^s_{\!(i+7)m} ] \;
[ {\cal D}^{s}_{\!2m} - i \, {\cal D}^s_{\!7m} ]
\nonumber \\[.2in] 
&\simeq& 
\frac{B_i \, \mu_j \, \tan\!\zb}{M_{\!s}^2} \,
\left[ \frac{\mu_{\scriptscriptstyle 0} 
M_{\!\scriptscriptstyle 1}  M_{\!\scriptscriptstyle 2} 
- M_{\!\scriptscriptstyle Z}^2 \, \sin\!\beta \cos\!\beta \,
 (M_{\!\scriptscriptstyle 1} \cos\!^2 \theta_{\!\scriptscriptstyle W}
 + M_{\!\scriptscriptstyle 2} \sin\!^2\theta_{\!\scriptscriptstyle W})}{\det({\cal M}_n)} \; 
\frac{1}{4} \,   (\tan\!\theta_{\!\scriptscriptstyle W} 
\mbox{\boldmath $X$}_{\!\!1n} - \mbox{\boldmath $X$}_{\!\!2n} )^2 \,
{M}_{\!\scriptscriptstyle \chi^0_{n}} \right]\;.
\eeqa \normalsize
This is exactly expression (\ref{N4j}) which we repeat.\\
\item
$\Pi_{\nu_i \!\tilde{h}}^{\scriptscriptstyle N} \, 
({\cal M}_n^{\mbox{-}1} \,  \xi^{\!\scriptscriptstyle  T})_{3j}$ ---
\footnotesize  \beqa
&& \mu_j \; \frac{M_{\!\scriptscriptstyle Z}^2  \cos^2\!\!\zb \; 
(M_{\!\scriptscriptstyle 1} \cos\!^2 \theta_{\!\scriptscriptstyle W}
 + M_{\!\scriptscriptstyle 2} \sin\!^2 \theta_{\!\scriptscriptstyle W})}{\det({\cal M}_n)} \;
\frac{1}{{4}} \,       [ \tan\!\theta_{\!\scriptscriptstyle W} 
\mbox{\boldmath $X$}_{\!\!1n} - \mbox{\boldmath $X$}_{\!\!2n} ]^2 \,
{M}_{\!\scriptscriptstyle \chi^0_{n}} \;
 \nonumber \\
&&  \mbox{\hspace*{2.5in} }\cdot
 [ {\cal D}^{s}_{\!(i+2)m} - i \, {\cal D}^s_{\!(i+7)m} ] \;
[ {\cal D}^{s}_{\!1m} - i \, {\cal D}^s_{\!6m} ]
\nonumber \\[.2in] 
&\simeq&
\frac{-B_i\, \mu_j \, \tan\!\zb}{M_s^2} \;
\left[ \frac{M_{\!\scriptscriptstyle Z}^2  \cos^2\!\!\zb \; 
(M_{\!\scriptscriptstyle 1} \cos\!^2 \theta_{\!\scriptscriptstyle W}
 + M_{\!\scriptscriptstyle 2} \sin\!^2 \theta_{\!\scriptscriptstyle W})}
{\det({\cal M}_n)} \;
\frac{1}{{4}} \,       ( \tan\!\theta_{\!\scriptscriptstyle W} 
\mbox{\boldmath $X$}_{\!\!1n} - \mbox{\boldmath $X$}_{\!\!2n} )^2 \,
{M}_{\!\scriptscriptstyle \chi^0_{n}} \right]\;.
\eeqa  \normalsize
Again, the $\Pi_{\nu_i \!\tilde{h}}^{\scriptscriptstyle N}$ term has a structure 
similar to that of $\Pi_{\nu_i 0}^{\scriptscriptstyle N}$ (or 
$\Pi_{\nu_{ij}}^{\scriptscriptstyle N}$) with the replacement of 
$\tilde{h}_{\!\scriptscriptstyle d}(\equiv {l}_{\scriptscriptstyle 0}$) by
$\tilde{h}_{\!\scriptscriptstyle u}^\dag$. \\
\item
$\Pi_{\nu_i \!\scriptscriptstyle  \tilde{W}}^{\scriptscriptstyle N}\, 
({\cal M}_n^{\mbox{-}1} \,  \xi^{\!\scriptscriptstyle  T})_{2j} $  --- part 1
\footnotesize  \beqa
& & \mu_j \; \frac{M_{\!\scriptscriptstyle Z} \cos\!\zb \;
\mu_{\scriptscriptstyle 0}  M_{\!\scriptscriptstyle 1} 
\cos\! \theta_{\!\scriptscriptstyle W}}{\det({\cal M}_n)} \;
\frac{1}{4}\, \mbox{\boldmath $X$}_{\!\!3n} \, [\tan\!\theta_{\!\scriptscriptstyle W} 
\mbox{\boldmath $X$}_{\!\!1n} - \mbox{\boldmath $X$}_{\!\!2n} ] \,
{M}_{\!\scriptscriptstyle \chi^0_{n}} \;
  [ {\cal D}^{s}_{\!(i+2)m} - i \, {\cal D}^s_{\!(i+7)m} ] \;
[ {\cal D}^{s}_{\!1m} + i \, {\cal D}^s_{\!6m} ] 
 \nonumber \\[.1in] 
& \simeq  &
\frac{-B_i\, \mu_j \, \tan\!\zb}{M_s^2} \;
\left[ \frac{M_{\!\scriptscriptstyle Z} \cos\!\zb \;
\mu_{\scriptscriptstyle 0}  M_{\!\scriptscriptstyle 1} 
\cos\! \theta_{\!\scriptscriptstyle W}}{\det({\cal M}_n)} \;
\frac{1}{4}\, \mbox{\boldmath $X$}_{\!\!3n} \, (\tan\!\theta_{\!\scriptscriptstyle W} 
\mbox{\boldmath $X$}_{\!\!1n} - \mbox{\boldmath $X$}_{\!\!2n} ) \,
{M}_{\!\scriptscriptstyle \chi^0_{n}} \right]\;.
\eeqa \normalsize
\\
$\Pi_{\nu_i \!\scriptscriptstyle  \tilde{W}}^{\scriptscriptstyle N}\, 
({\cal M}_n^{\mbox{-}1} \,  \xi^{\!\scriptscriptstyle  T})_{2j} $  --- part 2
\footnotesize  \beqa
& & - \mu_j \; \frac{M_{\!\scriptscriptstyle Z} \cos\!\zb \;
\mu_{\scriptscriptstyle 0}  M_{\!\scriptscriptstyle 1} 
\cos\! \theta_{\!\scriptscriptstyle W}}{\det({\cal M}_n)} \;
\frac{1}{4}\, \mbox{\boldmath $X$}_{\!\!(4+\alpha)n} \, [\tan\!\theta_{\!\scriptscriptstyle W} 
\mbox{\boldmath $X$}_{\!\!1n} - \mbox{\boldmath $X$}_{\!\!2n} ] \,
{M}_{\!\scriptscriptstyle \chi^0_{n}} \;
\nonumber \\
&&  \mbox{\hspace*{2in} }\cdot
  [ {\cal D}^{s}_{\!(i+2)m} - i \, {\cal D}^s_{\!(i+7)m} ] \;
[ {\cal D}^{s}_{\!(2+\alpha)m} + i \, {\cal D}^s_{\!(7+\alpha)m} ]
\nonumber \\[.2in]
& \simeq  &
\frac{B_i\, \mu_j \, \tan\!\zb}{M_s^2} \;
\left[ \frac{M_{\!\scriptscriptstyle Z} \cos\!\zb \;
\mu_{\scriptscriptstyle 0} M_{\!\scriptscriptstyle 1} 
\cos\! \theta_{\!\scriptscriptstyle W}}{\det({\cal M}_n)} \;
\frac{1}{4} \, \mbox{\boldmath $X$}_{\!\!4n} \,
 (\tan\!\theta_{\!\scriptscriptstyle W} 
\mbox{\boldmath $X$}_{\!\!1n} - \mbox{\boldmath $X$}_{\!\!2n} ) \,
{M}_{\!\scriptscriptstyle \chi^0_{n}} \right] \;.
\eeqa \normalsize
Here, we have different terms for $\za=0\mbox{-}3$, among which we show only the 
$\za=0$ result. The  $\za=1\mbox{-}3$ cases have obvious extra suppressions from
the $\mbox{\boldmath $X$}_{\!\!(4+\alpha)n}$ matrix element and a smaller scalar 
mixing part. The former has an extra  $\frac{\mu_j^*}{M_s}$ factor while the latter
introduces a   $\frac{(\tilde{m}_{\!\scriptscriptstyle {L}_{ij}}^2 + \mu_i^*\mu_j)}{M_s^2}$ 
factor. The overall $\Pi_{\nu_i  \!\scriptscriptstyle  \tilde{W}}^{\scriptscriptstyle N}$ 
results are not too different from the previous ones above either.
\item $\Pi_{\nu_i \!\scriptscriptstyle  \tilde{B}}^{\scriptscriptstyle N} \, 
({\cal M}_n^{\mbox{-}1} \,  \xi^{\!\scriptscriptstyle  T})_{1j}$ --- part 1
\footnotesize  \beqa
&& {- \mu_j } \; \frac{M_{\!\scriptscriptstyle Z} \cos\!\zb \;
\mu_{\scriptscriptstyle 0} M_{\!\scriptscriptstyle 2} 
\sin\! \theta_{\!\scriptscriptstyle W}}{\det({\cal M}_n)} \,
\frac{1}{4}\, \tan\!\theta_{\!\scriptscriptstyle W} 
\mbox{\boldmath $X$}_{\!\!3n} \, [\tan\!\theta_{\!\scriptscriptstyle W} 
\mbox{\boldmath $X$}_{\!\!1n} - \mbox{\boldmath $X$}_{\!\!2n} ] \,
{M}_{\!\scriptscriptstyle \chi^0_{n}} \;
  [ {\cal D}^{s}_{\!(i+2)m} - i \, {\cal D}^s_{\!(i+7)m} ] \;
[ {\cal D}^{s}_{\!1m} + i \, {\cal D}^s_{\!6m} ] 
\nonumber  \\[.1in]
&\simeq &
\frac{B_i\, \mu_j \, \tan\!\zb}{M_s^2} \;
\left[ \frac{M_{\!\scriptscriptstyle Z} \cos\!\zb \;
\mu_{\scriptscriptstyle 0} M_{\!\scriptscriptstyle 2} 
\sin\! \theta_{\!\scriptscriptstyle W}}{\det({\cal M}_n)} \,
\frac{1}{4}\, \tan\!\theta_{\!\scriptscriptstyle W} 
\mbox{\boldmath $X$}_{\!\!3n} \, (\tan\!\theta_{\!\scriptscriptstyle W} 
\mbox{\boldmath $X$}_{\!\!1n} - \mbox{\boldmath $X$}_{\!\!2n} ) \,
{M}_{\!\scriptscriptstyle \chi^0_{n}} \right]\;.
\eeqa \normalsize
 \\
$\Pi_{\nu_i \!\scriptscriptstyle  \tilde{B}}^{\scriptscriptstyle N} \, (
{\cal M}_n^{\mbox{-}1} \,  \xi^{\!\scriptscriptstyle  T})_{1j}$ --- part 2
\footnotesize  \beqa
&& - \mu_j \, \frac{M_{\!\scriptscriptstyle Z} \cos\!\zb \;
\mu_{\scriptscriptstyle 0} M_{\!\scriptscriptstyle 2} 
\sin\! \theta_{\!\scriptscriptstyle W}}{\det({\cal M}_n)} \,
\frac{1}{4}\, \tan\!\theta_{\!\scriptscriptstyle W} 
\mbox{\boldmath $X$}_{\!\!(4+\alpha)n} \, [\tan\!\theta_{\!\scriptscriptstyle W} 
\mbox{\boldmath $X$}_{\!\!1n} - \mbox{\boldmath $X$}_{\!\!2n} ] \,
{M}_{\!\scriptscriptstyle \chi^0_{n}} \;
\nonumber \\
&&  \mbox{\hspace*{2in} }\cdot
  [ {\cal D}^{s}_{\!(i+2)m} - i \, {\cal D}^s_{\!(i+7)m} ] \;
[ {\cal D}^{s}_{\!(2+\alpha)m} + i \, {\cal D}^s_{\!(7+\alpha)m} ]
\nonumber \\[.2in] 
&\simeq &
 \frac{B_i\, \mu_j \, \tan\!\zb}{M_s^2} \;
\left[ \frac{M_{\!\scriptscriptstyle Z} \cos\!\zb \;
\mu_{\scriptscriptstyle 0} M_{\!\scriptscriptstyle 2} 
\sin\! \theta_{\!\scriptscriptstyle W}}{\det({\cal M}_n)} \;
\frac{1}{4} \, \tan\!\theta_{\!\scriptscriptstyle W} 
\mbox{\boldmath $X$}_{\!\!4n} \,  (\tan\!\theta_{\!\scriptscriptstyle W} 
\mbox{\boldmath $X$}_{\!\!1n} - \mbox{\boldmath $X$}_{\!\!2n} ) \,
{M}_{\!\scriptscriptstyle \chi^0_{n}}  \right] \;.
\eeqa \normalsize
If one rotates the bino and wino into a photino and a zino, the photino would of 
course be decoupled from mass mixings with the neutral fermions. The closely 
related structures of  $\Pi_{\nu_i  \!\scriptscriptstyle  \tilde{W}}^{\scriptscriptstyle N} \, 
({\cal M}_n^{\mbox{-}1} \,  \xi^{\!\scriptscriptstyle  T})_{2j}$  
and $\Pi_{\nu_i  \!\scriptscriptstyle  \tilde{B}}^{\scriptscriptstyle N} \, 
({\cal M}_n^{\mbox{-}1} \,  \xi^{\!\scriptscriptstyle  T})_{1j}$ 
reflect on that. One can certainly write the two part of the results together
through a $\Pi_{\nu_i  \!\scriptscriptstyle  \tilde{Z}}^{\scriptscriptstyle N}$ 
term with diagrams involving the zino part only. However, to the extent that 
photino and zino are not mass eigenstates, there is really not much to gain. \\
\end{itemize}

\section{Details of \boldmath
$(\Pi_\xi \,  {\cal M}_n^{\mbox{-}1} \,  \xi^{\!\scriptscriptstyle  T})
^{\scriptscriptstyle C}$ Terms}

\begin{itemize}
\item $\Pi_{\nu_i \!0}^{\scriptscriptstyle C} \, ({\cal M}_n^{\mbox{-}1} \,
\xi^{\!\scriptscriptstyle T})_{4j} $ :-
\\
Here, we introduce the order 1 constant
\beq
C_4 = \frac{\mu_{\scriptscriptstyle 0} 
M_{\!\scriptscriptstyle 1}  M_{\!\scriptscriptstyle 2} 
- M_{\!\scriptscriptstyle Z}^2 \, \sin\!\beta \cos\!\beta \,
 (M_{\!\scriptscriptstyle 1} \cos\!^2 \theta_{\!\scriptscriptstyle W}
 + M_{\!\scriptscriptstyle 2} \sin\!^2\theta_{\!\scriptscriptstyle W})}
{\det({\cal M}_n)} \, M_s
\eeq
to simplify the expressions, given as follow.
\\
 $\Pi_{\nu_i \!0}^{\scriptscriptstyle C} \, ({\cal M}_n^{\mbox{-}1} \,
\xi^{\!\scriptscriptstyle T})_{4j} $ --- part 1
\beqa
\mu_j \; \frac{C_4}{M_s} \;
\frac{y_{\!\scriptscriptstyle e_i}}{g_{\scriptscriptstyle 2}} \,
\mbox{\boldmath $V$}^{*}_{\!\!(i+2)n} \, {M}_{\!\scriptscriptstyle \chi^{\mbox{-}}_{n}} \,
\mbox{\boldmath $U$}_{\!1n} \, D^{l}_{\!2m} \, D^{l^*}_{\!2m} 
\simeq 
\frac{y_{\!\scriptscriptstyle e_i}}{g_{\scriptscriptstyle 2}} \,
\frac{ m_i \,  \mu_i \,  \mu_j  }{M_{s}^2} \; C_4
\eeqa 
\\
$\Pi_{\nu_i \!0}^{\scriptscriptstyle C} \, ({\cal M}_n^{\mbox{-}1} \,
\xi^{\!\scriptscriptstyle T})_{4j} $ --- part 2
\beqa
\mu_j \;  \frac{C_4}{M_s} \;
\frac{y_{\!\scriptscriptstyle e_i}}{g_{\scriptscriptstyle 2}} \,
\frac{y_{\!\scriptscriptstyle e_k}}{g_{\scriptscriptstyle 2}} \;
\mbox{\boldmath $V$}^{*}_{\!\!(i+2)n} \, {M}_{\!\scriptscriptstyle \chi^{\mbox{-}}_{n}}  \,
\mbox{\boldmath $U$}_{\!(k+2)n} \, D^{l}_{\!2m} \, D^{l^*}_{\!(k+5)m}
\simeq  
\left( \frac{y_{\!\scriptscriptstyle e_i}}{g_{\scriptscriptstyle 2}} \right)^{\!\!2} 
\frac{m_i^2 \; \mu_i \, \mu_j  \tan\!\zb  }{M_{s}^3} \; C_4
\eeqa 
$\Pi_{\nu_i \!0}^{\scriptscriptstyle C} \, ({\cal M}_n^{\mbox{-}1} \,
\xi^{\!\scriptscriptstyle T})_{4j} $ --- part 3
\beqa
- \mu_j \;  \frac{C_4}{M_s} \; 
\frac{\lambda_{ikh}}{g_{\scriptscriptstyle 2}} \;
\mbox{\boldmath $V$}^{*}_{\!\!(h+2)n} \, {M}_{\!\scriptscriptstyle \chi^{\mbox{-}}_{n}} \,
 \mbox{\boldmath $U$}_{\!1n}  \,     D^{l}_{\!(k+2)m} \, D^{l^*}_{\!2m}
\simeq 
-\frac{\lambda_{\scriptscriptstyle ikh}}{g_{\scriptscriptstyle 2}} \,
\frac{\mu_j \, m_h \, \mu_h \, B_k^*\, \tan\!\zb}{M_{s}^4} \; C_4
\eeqa 
$\Pi_{\nu_i \!0}^{\scriptscriptstyle C} \, ({\cal M}_n^{\mbox{-}1} \,
\xi^{\!\scriptscriptstyle T})_{4j} $ --- part 4
\beqa
-\mu_j \,  \frac{C_4}{M_s} \,
\frac{\lambda_{ikh}}{g_{\scriptscriptstyle 2}} 
\frac{y_{\!\scriptscriptstyle e_{p}}}{g_{\scriptscriptstyle 2}}\,
\mbox{\boldmath $V$}^{*}_{\!\!(h+2)n}  {M}_{\!\scriptscriptstyle \chi^{\mbox{-}}_{n}} 
 \mbox{\boldmath $U$}_{\!(p+2)n}  \,     D^{l}_{\!(k+2)m}  D^{l^*}_{\!(p+5)m}
\simeq 
-\frac{\lambda_{ikh}}{g_{\scriptscriptstyle 2}} 
\frac{y_{\!\scriptscriptstyle e_h}}{g_{\scriptscriptstyle 2}}\,
\frac{\mu_j   m_h  (\widetilde{\cal M}_{\!\scriptscriptstyle RL}^{2})^*_{hk} }{M_{s}^3} \, C_4
\eeqa 
$\Pi_{\nu_i \!0}^{\scriptscriptstyle C} \, ({\cal M}_n^{\mbox{-}1} \,
 \xi^{\!\scriptscriptstyle T})_{4j} $ --- part 5
\beqa 
-\mu_j \; \frac{C_4}{M_s} \; 
\frac{y_{\!\scriptscriptstyle e_k}}{g_{\scriptscriptstyle 2}} \,
\mbox{\boldmath $V$}^*_{\!\!(k+2)n}  \, {M}_{\!\scriptscriptstyle \chi^{\mbox{-}}_{n}}  \,
\mbox{\boldmath $U$}_{\!1n}\, D^{l}_{\!(k+2)m} \, D^{l^*}_{\!(i+2)m}
\simeq 
-\frac{y_{\!\scriptscriptstyle e_i}}{g_{\scriptscriptstyle 2}} \,
\frac{m_i \, \mu_i \,  \mu_j \ }{M_{s}^2} \, C_4
\eeqa 
$\Pi_{\nu_i \!0}^{\scriptscriptstyle C} \, ({\cal M}_n^{\mbox{-}1} \,
\xi^{\!\scriptscriptstyle T})_{4j} $ --- part 6
\beqa 
\mu_j \;  \frac{C_4}{M_s} \; 
\frac{y_{\!\scriptscriptstyle e_i}}{g_{\scriptscriptstyle 2}} 
\frac{y_{\!\scriptscriptstyle e_k}}{g_{\scriptscriptstyle 2}} \,
\mbox{\boldmath $V$}^*_{\!\!(k+2)n}  \, {M}_{\!\scriptscriptstyle \chi^{\mbox{-}}_{n}}  \,
\mbox{\boldmath $U$}_{\!2n}\, D^{l}_{\!(k+2)m} \, D^{l^*}_{\!(i+5)m}
\simeq 
\frac{y_{\!\scriptscriptstyle e_i}}{g_{\scriptscriptstyle 2}} 
\frac{y_{\!\scriptscriptstyle e_k}}{g_{\scriptscriptstyle 2}} \,
\frac{\mu_j \,  m_k \,  \mu_k \,(\widetilde{\cal M}_{\!\scriptscriptstyle RL}^{2})^*_{ik}}{M_{s}^4} \,
C_4 
\eeqa 
$\Pi_{\nu_i \!0}^{\scriptscriptstyle C} \, ({\cal M}_n^{\mbox{-}1} \,
\xi^{\!\scriptscriptstyle T})_{4j} $ --- part 7
\beqa
-\mu_j \, \frac{C_4}{M_s} \,
\frac{\lambda_{ihk}}{g_{\scriptscriptstyle 2}} 
\frac{y_{\!\scriptscriptstyle e_p}}{g_{\scriptscriptstyle 2}}\,
\mbox{\boldmath $V$}^{*}_{\!\!(p+2)n} {M}_{\!\scriptscriptstyle \chi^{\mbox{-}}_{n}}  
\mbox{\boldmath $U$}_{\!(h+2)n} \, D^{l}_{\!(p+2)m} D^{l^*}_{\!(k+5)m} 
\simeq  
-\frac{\lambda_{ihk}}{g_{\scriptscriptstyle 2}} 
\frac{y_{\!\scriptscriptstyle e_h}}{g_{\scriptscriptstyle 2}}\,
\frac{\mu_j   m_h   (\widetilde{\cal M}_{\!\scriptscriptstyle RL}^{2})^*_{kh} }{M_{s}^3} \, C_4
\eeqa 
\item 
$\Pi_{\nu_i \!\tilde{h}}^{\scriptscriptstyle C} \, ({\cal M}_n^{\mbox{-}1} \,
\xi^{\!\scriptscriptstyle T})_{3j} $ :-
\\[.05in]
Here, we need to use, in addition to above, expressions for the elements of the 
mixing matrix  $D^{l}_{\!1(i+2)} \simeq \frac{B_i }{M_{s}^2}$
and $D^{l}_{\!1(i+5)} \simeq \frac{m_i \, \mu_i }{M_{s}^2}$; and 
introduce the order 1 constant
\beq
C_3 = \frac{M_{\!\scriptscriptstyle Z}^2  \cos^2\!\!\zb \; 
(M_{\!\scriptscriptstyle 1} \cos\!^2 \theta_{\!\scriptscriptstyle W}
 + M_{\!\scriptscriptstyle 2} \sin\!^2 \theta_{\!\scriptscriptstyle W})}{\det({\cal M}_n)} 
 \, {M}_{s} 
\eeq
to simplify the expressions. We also use $M_c$ to denote a mass parameter of the 
(physical) chargino mass scale. The results are as follow
\\[.05in]
$\Pi_{\nu_i \!\tilde{h}}^{\scriptscriptstyle C} \, ({\cal M}_n^{\mbox{-}1} \,
\xi^{\!\scriptscriptstyle T})_{3j} $ --- part 1
\beqa
& &
 \mu_j \; \frac{C_3}{M_{s}} \;
\mbox{\boldmath $V$}^{*}_{\!\!1n} \, {M}_{\!\scriptscriptstyle \chi^{\mbox{-}}_{n}} \,
\mbox{\boldmath $U$}_{\!1n} \,         D^{l}_{\!1m} \, D^{l^*}_{\!(i+2)m}
\simeq 
\frac{B_i\, \mu_j }{M_{s}^2} \; C_3 \frac{M_{c}}{M_{s}}
\eeqa 
$\Pi_{\nu_i \!\tilde{h}}^{\scriptscriptstyle C} \, ({\cal M}_n^{\mbox{-}1} \,
\xi^{\!\scriptscriptstyle T})_{3j} $ --- part 2
\beqa
& &
-\mu_j \;  \frac{C_3}{M_{s}} \;
   \frac{y_{\!\scriptscriptstyle e_i}}{g_{\scriptscriptstyle 2}} \,
\mbox{\boldmath $V$}^{*}_{\!\!1n} \, {M}_{\!\scriptscriptstyle \chi^{\mbox{-}}_{n}} \,
\mbox{\boldmath $U$}_{\!2n} \,         D^{l}_{\!1m} \, D^{l^*}_{\!(i+5)m}
\simeq 
-\frac{y_{\!\scriptscriptstyle e_i}}{g_{\scriptscriptstyle 2}} \,
\frac{m_i \, \mu_i \, \mu_j}{M_{s}^2}  \; C_3 \frac{M_{c}}{M_{s}}
\eeqa 
$\Pi_{\nu_i \!\tilde{h}}^{\scriptscriptstyle C} \, ({\cal M}_n^{\mbox{-}1} \,
\xi^{\!\scriptscriptstyle T})_{3j} $ --- part 3
\beqa
& &
\mu_j \;  \frac{C_3}{M_{s}} \;
 \frac{ \lambda_{ihk}}{g_{\scriptscriptstyle 2}}  \, 
\mbox{\boldmath $V$}^{*}_{\!\!1n} \, {M}_{\!\scriptscriptstyle \chi^{\mbox{-}}_{n}} \,
\mbox{\boldmath $U$}_{\!(h+2)n} \,         D^{l}_{\!1m} \, D^{l^*}_{\!(k+5)m}
\simeq  
\frac{ \lambda_{ihk}} { g_{\scriptscriptstyle 2}} \,
\frac{\mu_j \, m_h \, \mu_h^* \, m_k \, \mu_k} {M_{s}^4} \, C_3 \,
\eeqa 
\item 
$\Pi_{\nu_i \!\tilde{W}}^{\scriptscriptstyle C} \, ({\cal M}_n^{\mbox{-}1} \,
\xi^{\!\scriptscriptstyle T})_{2j} $ and 
$\Pi_{\nu_i \!\tilde{B}}^{\scriptscriptstyle C} \, ({\cal M}_n^{\mbox{-}1} \,
\xi^{\!\scriptscriptstyle T})_{1j} $ :- 
\\[.1in]
We have noted above in the case of $\Pi_{\xi}^{\scriptscriptstyle N}$ the close similarity
between $\Pi_{\nu_i \!\tilde{W}}^{\scriptscriptstyle N} \, ({\cal M}_n^{\mbox{-}1} \,
\xi^{\!\scriptscriptstyle T})_{2j} $ and 
$\Pi_{\\nu_i \!\tilde{B}}^{\scriptscriptstyle N} \, ({\cal M}_n^{\mbox{-}1} \,
\xi^{\!\scriptscriptstyle T})_{1j} $. The story in the same here, between 
$\Pi_{\nu_i \!\tilde{W}}^{\scriptscriptstyle C} \, ({\cal M}_n^{\mbox{-}1} \,
\xi^{\!\scriptscriptstyle T})_{2j} $ and 
$\Pi_{\nu_i \!\tilde{B}}^{\scriptscriptstyle C} \, ({\cal M}_n^{\mbox{-}1} \,
\xi^{\!\scriptscriptstyle T})_{1j} $, with some exception.
Note that from comparing Eqs.(\ref{CWnm}) and (\ref{CBnm}), we can see that there
is and extra term in 
${\cal C}^{\scriptscriptstyle R}_{\!\scriptscriptstyle \tilde{B}\!nm}$
without the matching partner in 
${\cal C}^{\scriptscriptstyle R}_{\!\scriptscriptstyle \tilde{W}\!nm}$. 
We list below all results of the bino case, namely,
$\Pi_{\nu_i \!\tilde{B}}^{\scriptscriptstyle C} \, ({\cal M}_n^{\mbox{-}1} \,
\xi^{\!\scriptscriptstyle T})_{1j}$. Among the ten parts listed below, 1 to 7
have the wino counterparts in $\Pi_{\nu_i \!\tilde{B}}^{\scriptscriptstyle C} \, 
({\cal M}_n^{\mbox{-}1} \, \xi^{\!\scriptscriptstyle T})_{2j}$, to be given
with an extra factor $\frac{-1}{\tan\!\theta_{\!\scriptscriptstyle W} }
\frac{M_{\!\scriptscriptstyle 1}}{M_{\!\scriptscriptstyle 2}}$, which
we do not list explicitly. Parts 8 to 10 have no wino counterparts.
We also introduce order 1 constants
\beqa
C_1 &=& \frac{\tan\!\theta_{\!\scriptscriptstyle W} }{\sqrt{2}} \,
\frac{M_{\!\scriptscriptstyle Z} \cos\!\zb \;
\mu_{\scriptscriptstyle 0} M_{\!\scriptscriptstyle 2} 
\sin\! \theta_{\!\scriptscriptstyle W}}{\det({\cal M}_n)} \, M_{s}
\eeqa
to simplify the expressions. 
\\[.15in]
$\Pi_{\nu_i \!\tilde{B}}^{\scriptscriptstyle C} \, ({\cal M}_n^{\mbox{-}1} \,
\xi^{\!\scriptscriptstyle T})_{1j} $ --- part 1
\beqa
& &  {- \mu_j } \; \frac{C_1}{M_{s}} \;
\frac{y_{\!\scriptscriptstyle e_i}}{g_{\scriptscriptstyle 2}} \,
\mbox{\boldmath $V$}^*_{\!\!(i+2)n} \, {M}_{\!\scriptscriptstyle \chi^{\mbox{-}}_{n}} \, 
\mbox{\boldmath $U$}_{\!2n}                 \, D^l_{\!2m} \, D^{l^*}_{\!2m}
\simeq 
-\frac{y_{\!\scriptscriptstyle e_i}}{g_{\scriptscriptstyle 2}} \,
 \frac{ m_i \, \mu_i \,  \mu_j }{M_{s}^2} \,
C_1 
\eeqa 
$\Pi_{\nu_i \!\tilde{B}}^{\scriptscriptstyle C} \, ({\cal M}_n^{\mbox{-}1} \,
\xi^{\!\scriptscriptstyle T})_{1j} $ --- part 2
\beq
 {- \mu_j } \;  \frac{C_1}{M_{s}} \;
\frac{y_{\!\scriptscriptstyle e_i}}{g_{\scriptscriptstyle 2}} \,
\mbox{\boldmath $V$}^{*}_{\!\!(i+2)n} \, {M}_{\!\scriptscriptstyle \chi^{\mbox{-}}_{n}}  \,
\mbox{\boldmath $U$}_{\!(k+2)n} \, D^l_{\!2m} \, D^{l^*}_{\!(k+2)m}
\simeq 
-\frac{y_{\!\scriptscriptstyle e_i}}{g_{\scriptscriptstyle 2}} \;
\frac{m_i \,  B_i \,  \mu_j \, \tan\!\zb }{M^3_{\!s}} \, 
C_1 
\eeq 
$\Pi_{\nu_i \!\tilde{B}}^{\scriptscriptstyle C} \, ({\cal M}_n^{\mbox{-}1} \,
\xi^{\!\scriptscriptstyle T})_{1j} $ --- part 3
\beqa
& & { \mu_j } \;  \frac{C_1}{M_{s}} \;
\frac{\lambda_{ikh}}{g_{\scriptscriptstyle 2}} \,
\mbox{\boldmath $V$}^*_{\!\!(h+2)n} \, {M}_{\!\scriptscriptstyle \chi^{\mbox{-}}_{n}} \,
 \mbox{\boldmath $U$}_{\!2n} \,          D^l_{\!(k+2)m} \, D^{l^*}_{\!2m}
\simeq
\frac{\lambda_{ikh}}{g_{\scriptscriptstyle 2}} \,
\frac{ \mu_j \, m_h \, \mu_h  \, B^*_k \, \tan\!\zb}{M_{s}^4} \, 
C_1
\eeqa 
$\Pi_{\nu_i \!\tilde{B}}^{\scriptscriptstyle C} \, ({\cal M}_n^{\mbox{-}1} \,
\xi^{\!\scriptscriptstyle T})_{1j} $ --- part 4
\beqa
 & &{ \mu_j } \;  \frac{C_1}{M_{s}} \;
\frac{\lambda_{ikh}}{g_{\scriptscriptstyle 2}} \,
\mbox{\boldmath $V$}^*_{\!\!(h+2)n} \, {M}_{\!\scriptscriptstyle \chi^{\mbox{-}}_{n}} \,
 \mbox{\boldmath $U$}_{\!(p+2)n} \,
     D^l_{\!(k+2)m}  \, D^{l^*}_{\!(p+2)m} \
\simeq 
\frac{\lambda_{ikh}}{g_{\scriptscriptstyle 2}} \;
\frac{\mu_j \, m_h \,  (\tilde{m}_{\!\scriptscriptstyle {L}_{kh}}^2 + \mu_k^*\mu_h) }{M_{s}^3} \,
C_1 
\eeqa 
$\Pi_{\nu_i \!\tilde{B}}^{\scriptscriptstyle C} \, ({\cal M}_n^{\mbox{-}1} \,
\xi^{\!\scriptscriptstyle T})_{1j} $ --- part 5
\beqa
& & {- \mu_j } \; \frac{C_1}{M_{s}} \;
\mbox{\boldmath $V$}^*_{\!\!2n}  \, {M}_{\!\scriptscriptstyle \chi^{\mbox{-}}_{n}} \,
\mbox{\boldmath $U$}_{\!1n}\,         D^l_{\!1m} \,D^{l^*}_{\!(i+2)m} \, 
\simeq 
-\frac{B_i \, \mu_j \ }{M_{s}^2} \,
C_1  \frac{M_{c}}{M_{s}}
\eeqa 
$\Pi_{\nu_i \!\tilde{B}}^{\scriptscriptstyle C} \, ({\cal M}_n^{\mbox{-}1} \,
\xi^{\!\scriptscriptstyle T})_{1j} $ --- part 6
\beqa
& & { \mu_j } \;  \frac{C_1}{M_{s}} \;
\frac{y_{\!\scriptscriptstyle e_i}}{g_{\scriptscriptstyle 2}} \,
\mbox{\boldmath $V$}^*_{\!\!2n} \, {M}_{\!\scriptscriptstyle \chi^{\mbox{-}}_{n}} \, 
\mbox{\boldmath $U$}_{\!2n} \,       D^l_{\!1m} \, D^{l^*}_{\!(i+5)m} 
\simeq 
\frac{y_{\!\scriptscriptstyle e_i}}{g_{\scriptscriptstyle 2}} \,
\frac{m_i \,  \mu_i \, \mu_j \ }{M_{s}^2} \,
C_1  \frac{M_{c}}{M_{s}}
\eeqa 
$\Pi_{\nu_i \!\tilde{B}}^{\scriptscriptstyle C} \, ({\cal M}_n^{\mbox{-}1} \,
\xi^{\!\scriptscriptstyle T})_{1j} $ --- part 7
\beqa
& & {- \mu_j } \; \frac{C_1}{M_{s}} \;
\frac{\lambda_{ihk}}{g_{\scriptscriptstyle 2}} \,
\mbox{\boldmath $V$}^*_{\!\!2n} \, {M}_{\!\scriptscriptstyle \chi^{\mbox{-}}_{n}} \,
\mbox{\boldmath $U$}_{\!(h+2)n}  \,            D^l_{\!1m} \, D^{l^*}_{\!(k+5)m} \ 
\simeq  
-\frac{\lambda_{ihk}}{g_{\scriptscriptstyle 2}} \,
\frac{\mu_j \, \mu_h^* \,  \mu_k \, m_k  }{M_{s}^3} \, 
C_1
\eeqa 
$\Pi_{\nu_i \!\tilde{B}}^{\scriptscriptstyle C} \, ({\cal M}_n^{\mbox{-}1} \,
\xi^{\!\scriptscriptstyle T})_{1j} $ --- part 8
\beqa
& & { -\mu_j } \; \frac{C_1}{M_{s}} \;2\, 
\mbox{\boldmath $V$}^*_{\!\!(k+2)n}\, {M}_{\!\scriptscriptstyle \chi^{\mbox{-}}_{n}} \,
\mbox{\boldmath $U$}_{\!1n}  \,          D^l_{\!(k+5)m} \, D^{l^*}_{\!(i+2)m}
\simeq 
\frac{\mu_j \, m_k \,  \mu_{k} \, 
(\widetilde{\cal M}_{\!\scriptscriptstyle RL}^{2})^*_{ik}}{M_{s}^4} \,  
2\, C_1
\eeqa 
$\Pi_{\nu_i \!\tilde{B}}^{\scriptscriptstyle C} \, ({\cal M}_n^{\mbox{-}1} \,
\xi^{\!\scriptscriptstyle T})_{1j} $ --- part 9
\beqa
& & { \mu_j } \; \frac{C_1}{M_{s}} \;2\, 
\frac{y_{\!\scriptscriptstyle e_i}}{g_{\scriptscriptstyle 2}} \,
\mbox{\boldmath $V$}^*_{\!\!(k+2)n}  \, {M}_{\!\scriptscriptstyle \chi^{\mbox{-}}_{n}} \, 
 \mbox{\boldmath $U$}_{\!2n}\,       D^l_{\!(k+5)m} \, D^{l^*}_{\!(i+5)m} \
\simeq 
\frac{y_{\!\scriptscriptstyle e_i}}{g_{\scriptscriptstyle 2}} \;
\frac{ m_i \,  \mu_i \,  \mu_j }{M_{s}^2}  \, 
2\, C_1
\eeqa 
$\Pi_{\nu_i \!\tilde{B}}^{\scriptscriptstyle C} \, ({\cal M}_n^{\mbox{-}1} \,
\xi^{\!\scriptscriptstyle T})_{1j} $ --- part 10
\beqa
& & {- \mu_j } \; \frac{C_1}{M_{s}} \;2\, 
\frac{\lambda_{ihk}}{g_{\scriptscriptstyle 2}} \,
\mbox{\boldmath $V$}^*_{\!\!(p+2)n}  \, {M}_{\!\scriptscriptstyle \chi^{\mbox{-}}_{n}} \, 
\mbox{\boldmath $U$}_{\!(h+2)n}\,            D^l_{\!(p+5)m} \, D^{l^*}_{\!(k+5)m}\
 \simeq 
-\frac{\lambda_{ihk}}{g_{\scriptscriptstyle 2}} \,
\frac{ \mu_j \, m_k \, \tilde{m}_{\!\scriptscriptstyle {E}_{hk}}^2 }{M_{s}^3}  \, 
2\, C_1
\eeqa 
\\
\end{itemize}

\section{Details of \boldmath
$(\Pi_\xi \,  {\cal M}_n^{\mbox{-}1} \,  \xi^{\!\scriptscriptstyle  T})
^{D}$ Terms}
\noindent
{\bf\boldmath $\Pi_{\nu_{ij}}^{\scriptscriptstyle D}$ :-} \\[.2in]
 Note that the necessary symmetrization is not shown explicitly.
\begin{itemize}
\item ~ \vspace*{-.3in}
\beqa
{N_c}\,{ \lambda_{ikh}^{\prime}  \,\over g_{\scriptscriptstyle 2}} \,
{ \lambda_{jhp}^{\prime}  \,\over g_{\scriptscriptstyle 2}} \;{m}_{\!\scriptscriptstyle d_{h}} \; 
 {\cal D}^{d}_{\!km} \, {\cal D}^{d^*}_{\!(p+3)m} 
\simeq  
3\, { \lambda_{ ikh}^{\prime} \, \over  g_{\scriptscriptstyle 2} } \,
{ \lambda_{jhp}^{\prime} \,\over g_{\scriptscriptstyle 2} }  \;
{ m_{ \scriptscriptstyle d_{h} } \, 
({\cal M}_{\!\scriptscriptstyle RL}^{2})^*_{pk} \over M^2_{s}} \; ,
\eeqa
where
\beqa
({\cal M}_{\!\scriptscriptstyle RL}^{2})_{pk} 
=  [A_{\scriptscriptstyle d}^* - \mu_{\scriptscriptstyle 0} \, \tan\!\beta ] \, 
m_{\scriptscriptstyle d_{p}} \, \delta_{kp}
+ \frac{\sqrt{2}\, M_{\!\scriptscriptstyle W} \cos\!\beta}
{g_{\scriptscriptstyle 2} } \,   \delta\! A^{\!{\scriptscriptstyle D}^*}_{kp}
 - \frac{\sqrt{2}\, M_{\!\scriptscriptstyle W} \sin\!\beta}
{g_{\scriptscriptstyle 2} } \,(\, \mu_i\lambda_{ikp}^{\prime \,*} \, )  \; .
\eeqa
\end{itemize}

\noindent
{\bf\boldmath
$(\Pi_\xi \,  {\cal M}_n^{\mbox{-}1} \,  \xi^{\!\scriptscriptstyle  T})
^{\scriptscriptstyle D}_{ij}$ :-} \\[.2in]
$\Pi_{\nu_i \!\tilde{0}}^{\scriptscriptstyle D} \, ({\cal M}_n^{\mbox{-}1} \,
\xi^{\!\scriptscriptstyle T})_{4j} $ ---part 1
\beqa
-{N_c}\,\mu_j \, \frac{C_4}{M_{s}}\,
\frac{\lambda^{\prime}_{ikh} \,}{g_{\scriptscriptstyle 2}^2} \, 
\frac{ y_{\!\scriptscriptstyle d_h} \,} {g_{\scriptscriptstyle 2}^2} \, 
m_{\scriptscriptstyle d_h} \, {\cal D}^{d}_{\!km} \, {\cal D}^{d^*}_{\!(h+3)m} 
\simeq 
-\frac{\lambda_{ikh}^{\prime} \,}{g_{\scriptscriptstyle 2}} \, 
  \frac{y_{\!\scriptscriptstyle d_h} \,}{g_{\scriptscriptstyle 2}} \, 
 \frac{\mu_j \, m_{\scriptscriptstyle d_h} \,
({\cal M}_{\!\scriptscriptstyle RL}^{2})^*_{hk} \, }{M_{s}^3} \, 3C_4
\eeqa 
\\
$\Pi_{\nu_i \!\tilde{0}}^{\scriptscriptstyle D} \, ({\cal M}_n^{\mbox{-}1} \,
\xi^{\!\scriptscriptstyle T})_{4j} $ ---part 2
\beqa
-{N_c}\,\mu_j \, \frac{C_4}{M_{s}}\,
\frac{\lambda^{\prime}_{ihk} \,}{g_{\scriptscriptstyle 2}} \, 
\frac{ y_{\!\scriptscriptstyle d_h} \,} {g_{\scriptscriptstyle 2}} \, 
m_{\scriptscriptstyle d_h} \, {\cal D}^{d}_{\!hm} \, {\cal D}^{d^*}_{\!(k+3)m}     
\simeq 
-\frac{\lambda_{ihk}^{\prime} \,}{g_{\scriptscriptstyle 2}} \, 
  \frac{y_{\!\scriptscriptstyle d_h} \,}{g_{\scriptscriptstyle 2}} \, 
 \frac{\mu_j \, m_{\scriptscriptstyle d_n} \,
({\cal M}_{\!\scriptscriptstyle RL}^{2})^*_{kh} \, }{M_{s}^3} \, 3C_4
\eeqa 
\\
$\Pi_{\nu_i \!\tilde{W}}^{\scriptscriptstyle D} \, ({\cal M}_n^{\mbox{-}1} \,
\xi^{\!\scriptscriptstyle T})_{2j} $ 
\beqa
{N_c}\, { \mu_j } \, \frac{C_1}{\tan\!\theta_{\!\scriptscriptstyle W}\, M_s}
\frac{\lambda_{ikh}^{\prime} \, } {g_{\scriptscriptstyle 2}} \,
m_{\scriptscriptstyle d_h} \, {\cal D}^{d}_{\!km} \, {\cal D}^{d^*}_{\!hm} 
\simeq  
\frac{\lambda_{ ikh}^{\prime} \,} {g_{\scriptscriptstyle 2}} \, 
\frac{\mu_j \, m_{\scriptscriptstyle d_h} \,\tilde{m}^2_{\!\scriptscriptstyle Q_{kh}}}{M_s^3}\; 
\frac{3C_1}{ \tan\!\theta_{\!\scriptscriptstyle W}}\,
\eeqa
\\ 
$\Pi_{\nu_i \!\tilde{B}}^{\scriptscriptstyle D} \, ({\cal M}_n^{\mbox{-}1} \,
\xi^{\!\scriptscriptstyle T})_{1j} $ --- part 1
\beqa
-{N_c}\, { \mu_j } \, \frac{C_1}{3 \, M_s}
\frac{\lambda_{ikh}^{\prime} \,}{g_{\scriptscriptstyle 2}} \,
 m_{\scriptscriptstyle d_h} \, {\cal D}^{d}_{\!km} \, {\cal D}^{d^*}_{\!hm} 
\simeq 
-\frac{\lambda_{ikh}^{\prime} \, }{g_{\scriptscriptstyle 2}} \,
\frac{ \mu_j \, m_{\scriptscriptstyle d_h} \,\tilde{m}^2_{\!\scriptscriptstyle Q_{kh}}}{M_{s}^3} \, C_1
\eeqa
\\
$\Pi_{\nu_i \!\tilde{B}}^{\scriptscriptstyle D} \, ({\cal M}_n^{\mbox{-}1} \,
\xi^{\!\scriptscriptstyle T})_{1j} $ --- part 2
\beqa
-{N_c}\,{ \mu_j } \, \frac{2 \, C_1}{3\, M_{s}}
\frac{\lambda_{ihk}^{\prime} \, }{g_{\scriptscriptstyle 2}} \,m_{\scriptscriptstyle d_h} \,
           {\cal D}^{d}_{\!(h+3)m} \, {\cal D}^{d^*}_{\!(k+3)m}  
\simeq 
-\frac{\lambda_{ihk}^{\prime} \,} {g_{\scriptscriptstyle 2}} \, 
\frac{ \mu_j \, m_{\scriptscriptstyle d_h}\, \tilde{m}^2_{\!\scriptscriptstyle D_{hk}}}{M_{s}^3} \; 2C_1
\eeqa

\section{The $\Sigma_\nu$ Results}
\noindent
{\bf\boldmath $\Sigma_{\nu_{ij}}^{\scriptscriptstyle N}$ :-} \\[.1in]
We have a simple result here, given as
\begin{itemize}
\item ~ \vspace*{-.3in}
\beqa
&&
\frac{1}{{4}} \,       
[ \tan\!\theta_{\!\scriptscriptstyle W} 
\mbox{\boldmath $X$}_{\!\!1n} - \mbox{\boldmath $X$}_{\!\!2n} ]^2 \,
  [ {\cal D}^{s}_{\!(i+2)m} - i \, {\cal D}^s_{\!(i+7)m} ] \;
[ {\cal D}^{s}_{\!(j+2)m} + i \, {\cal D}^s_{\!(j+7)m} ]
\nonumber \\ 
&\simeq&
\frac{B_i B_j^*\, \tan\!^2\!\zb }{M_{s}^4} \; 
\left[ \frac{1}{{4}} \,  (\tan\!\theta_{\!\scriptscriptstyle W} 
\mbox{\boldmath $X$}_{\!\!1n} - \mbox{\boldmath $X$}_{\!\!2n} )^2 \,
\right] \,
 \nonumber \\ 
&&\mbox{\hspace*{4in} (n=1-4 dominate)} \;.
\eeqa
\end{itemize}
\noindent
{\bf\boldmath $\Sigma_{\nu_{ij}}^{\scriptscriptstyle C}$ :-} \\[.1in]
We list all the individual terms below.
\begin{itemize}
\item ~ \vspace*{-.3in}
\beq
\mbox{\boldmath $U$}_{\!\!1n} \, \mbox{\boldmath $U$}_{\!\!1n}^*\,
  D^{l^*}_{\!(i+2)m}\,  D^{l}_{\!(j+2)m} \;  
\simeq  \frac{ (\tilde{m}^2_{\!\scriptscriptstyle L_{ji}}+\mu_{i} \,  \mu_{j}^*)}{M_{s}^2}
\eeq  
\item ~ \vspace*{-.3in}
\beq
 \frac{y_{\!\scriptscriptstyle e_i}}{g_{\scriptscriptstyle 2}} \, 
       \frac{y_{\!\scriptscriptstyle e_j}}{g_{\scriptscriptstyle 2}} \, 
       \mbox{\boldmath $U$}_{\!\!2n} \, \mbox{\boldmath $U$}_{\!\!2n}^* \,
        D^{l^*}_{\!(i+5)m} \,  D^{l}_{\!(j+5)m} \;  \simeq 
       \frac{y_{\!\scriptscriptstyle e_i}}{g_{\scriptscriptstyle 2}} \, 
       \frac{y_{\!\scriptscriptstyle e_j}}{g_{\scriptscriptstyle 2}} \,
      \frac{\tilde{m}_{\!\scriptscriptstyle E_{ji}}^2}{M_{s}^2}
\eeq  
\item ~ \vspace*{-.3in}
\beq
{\lambda_{ihk} \over g_{\scriptscriptstyle 2} } \, 
      {\lambda_{jpq}^* \over g_{\scriptscriptstyle 2} } \, 
      \mbox{\boldmath $U$}_{\!\!(h+2)n} \, 
        \mbox{\boldmath $U$}_{\!\!(p+2)n}^* \,
        D^{l^*}_{\!(k+5)m} \, D^{l}_{\!(q+5)m} \;  \simeq
      {\lambda_{ihk} \over g_{\scriptscriptstyle 2} } \, 
      {\lambda_{jhk}^* \over g_{\scriptscriptstyle 2} } \;
\eeq  
\item ~ \vspace*{-.3in}
\beq
-\frac{y_{\!\scriptscriptstyle e_j}}{g_{\scriptscriptstyle 2}} \, 
      \mbox{\boldmath $U$}_{\!\!1n} \, \mbox{\boldmath $U$}_{\!\!2n}^* \,
        D^{l^*}_{\!(i+2)m} \,  D^{l}_{\!(j+5)m} \;  \simeq
        -\frac{y_{\!\scriptscriptstyle e_j}}{g_{\scriptscriptstyle 2}} \,
       \frac{(\widetilde{M}_{\!\scriptscriptstyle RL}^2)_{ji}}{M_s^2}  \; 
\eeq  
\item ~ \vspace*{-.3in}
\beq
-\frac{y_{\!\scriptscriptstyle e_j}}{g_{\scriptscriptstyle 2}} \, 
      \mbox{\boldmath $U$}_{\!\!2n} \, \mbox{\boldmath $U$}_{\!\!1n}^* \,
        D^{l^*}_{\!(i+5)m} \,  D^{l}_{\!(j+2)m} \;  \simeq
        -\frac{y_{\!\scriptscriptstyle e_i}}{g_{\scriptscriptstyle 2}} \,
       \frac{(\widetilde{M}_{\!\scriptscriptstyle RL}^2)^*_{ij}}{M_s^2}  \;
\eeq  
\item ~ \vspace*{-.3in}
\beq
{\lambda_{jhk}^* \over g_{\scriptscriptstyle 2} } \,
 \mbox{\boldmath $U$}_{\!\!1n} \, \mbox{\boldmath $U$}_{\!\!(h+2)n}^* \,
        D^{l^*}_{\!(i+2)m} \;  D^{l}_{\!(k+5)m} \;  \simeq
       {\lambda_{jhk}^* \over g_{\scriptscriptstyle 2} } \, 
      \frac{\mu_h\, (\widetilde{M}_{\!\scriptscriptstyle RL}^2)_{ki}}{M_s^3}
\eeq  
\item ~ \vspace*{-.3in}
\beq
{\lambda_{ihk} \over g_{\scriptscriptstyle 2} } \,
 \mbox{\boldmath $U$}_{\!\!(h+2)n} \, \mbox{\boldmath $U$}_{\!\!1n}^* \,
        D^{l^*}_{\!(k+5)m} \;  D^{l}_{\!(j+2)m} \;  \simeq
       {\lambda_{ihk} \over g_{\scriptscriptstyle 2} } \, 
      \frac{\mu_h^* \, (\widetilde{M}_{\!\scriptscriptstyle RL}^2)^*_{kj}}{M_s^3}
\eeq  
\item ~ \vspace*{-.3in}
\beq
-\frac{y_{\!\scriptscriptstyle e_i}}{g_{\scriptscriptstyle 2}} \,
      {\lambda_{jhk}^* \over g_{\scriptscriptstyle 2} } \,
      \mbox{\boldmath $U$}_{\!\!2n} \, \mbox{\boldmath $U$}_{\!\!(h+2)n}^* \,
        D^{l^*}_{\!(i+5)m} \;  D^{l}_{\!(k+5)m} \;  \simeq
      \frac{y_{\!\scriptscriptstyle e_i}}{g_{\scriptscriptstyle 2}} \,
       {\lambda_{jhi}^* \over g_{\scriptscriptstyle 2} } \, 
      \frac{\mu_h}{M_s} \; 
\eeq  
\item ~ \vspace*{-.3in}
\beq
-\frac{y_{\!\scriptscriptstyle e_j}}{g_{\scriptscriptstyle 2}} \,
      {\lambda_{ihk} \over g_{\scriptscriptstyle 2} } \,
      \mbox{\boldmath $U$}_{\!\!(h+2)n} \, \mbox{\boldmath $U$}_{\!\!2n}^* \,
        D^{l^*}_{\!(k+5)m} \;  D^{l}_{\!(j+5)m} \;  \simeq
      \frac{y_{\!\scriptscriptstyle e_j}}{g_{\scriptscriptstyle 2}} \,
      {\lambda_{ihj} \over g_{\scriptscriptstyle 2} } \, 
      \frac{\mu_h^*}{M_s} \; 
\eeq  
\item ~ \vspace*{-.3in}
\beq
\frac{y_{\!\scriptscriptstyle e_i}}{g_{\scriptscriptstyle 2}} \, 
        \frac{y_{\!\scriptscriptstyle e_j}}{g_{\scriptscriptstyle 2}} \, 
       \mbox{\boldmath $V$}_{\!\!(i+2)n}^* \, \mbox{\boldmath $V$}_{\!\!(j+2)n} \,
        D^{l}_{\!2m} \,  D^{l^*}_{\!2m} \;  \simeq 
       \frac{y_{\!\scriptscriptstyle e_i}}{g_{\scriptscriptstyle 2}} \, 
       \frac{y_{\!\scriptscriptstyle e_j}}{g_{\scriptscriptstyle 2}} \,
       \delta_{ij}
\eeq  
\item ~ \vspace*{-.3in}
\beq
 {\lambda_{ihk}   \over g_{\scriptscriptstyle 2} } \, 
        {\lambda_{jpq}^* \over g_{\scriptscriptstyle 2} } \, 
      \mbox{\boldmath $V$}_{\!\!(h+2)n}^* \, \mbox{\boldmath $V$}_{\!\!(p+2)n} \,
        D^{l}_{\!(k+2)m} \, D^{l^*}_{\!(q+2)m} \;  \simeq
      {\lambda_{ihk} \over g_{\scriptscriptstyle 2} } \, 
      {\lambda_{jhq}^* \over g_{\scriptscriptstyle 2} } \; 
      \frac{\tilde{m}^2_{\!\scriptscriptstyle L_{kq}}+\mu^*_{k}\,\mu_{q}}{M_s^2}
\eeq  
\item ~ \vspace*{-.3in}
\beq
-\frac{y_{\!\scriptscriptstyle e_i}}{g_{\scriptscriptstyle 2}} \,
      {\lambda_{jkh}^* \over g_{\scriptscriptstyle 2} } \,
      \mbox{\boldmath $V$}_{\!\!(i+2)n}^* \, \mbox{\boldmath $V$}_{\!\!(h+2)n} \,
        D^{l^*}_{\!(k+2)m} \,  D^{l}_{\!2m} \;  \simeq
      -\frac{y_{\!\scriptscriptstyle e_i}}{g_{\scriptscriptstyle 2}} \,
      {\lambda_{jki}^* \over g_{\scriptscriptstyle 2} } \, 
      \frac{B_{k} \, \tan{\!\beta} }{M_s^2} 
\eeq  
\item ~ \vspace*{-.3in}
\beq
-\frac{y_{\!\scriptscriptstyle e_j}}{g_{\scriptscriptstyle 2}} \,
      {\lambda_{ikh} \over g_{\scriptscriptstyle 2} } \,
      \mbox{\boldmath $V$}_{\!\!(j+2)n} \, \mbox{\boldmath $V$}_{\!\!(h+2)n}^* \,
        D^{l}_{\!(k+2)m}  \, D^{l^*}_{\!2m} \;  \simeq
      -\frac{y_{\!\scriptscriptstyle e_j}}{g_{\scriptscriptstyle 2}} \,
      {\lambda_{ikj} \over g_{\scriptscriptstyle 2} } \, 
      \frac{B_{k}^* \, \tan{\!\beta} }{M_s^2}
\eeq
\end{itemize}
\noindent
{\bf\boldmath $\Sigma_{\nu_{ij}}^{\scriptscriptstyle D}$ :-} 
\begin{itemize}
\item ~ \vspace*{-.3in}
\beq
N_{c} \, 
        {\lambda_{ink}^{\prime^*}  \over g_{\scriptscriptstyle 2} } \, 
        {\lambda_{jnq}^{\prime} \over g_{\scriptscriptstyle 2} } \, 
        D^{*}_{\!d_k2m} \,  D_{\!d_q2m} \;  \simeq
        3\, {\lambda_{ink}^{\prime^*} \over g_{\scriptscriptstyle 2} } \, 
      {\lambda_{jnq}^{\prime} \over g_{\scriptscriptstyle 2} } \; 
\eeq  
\item ~ \vspace*{-.3in}
\beq
N_{c} \, 
        {\lambda_{ikn}^{\prime^*}  \over g_{\scriptscriptstyle 2} } \, 
        {\lambda_{jqn}^{\prime} \over g_{\scriptscriptstyle 2} } \, 
        D^{*}_{\!d_k1m} \,  D_{\!d_q1m} \;  \simeq
        3\,       {\lambda_{ikn}^{\prime^*} \over g_{\scriptscriptstyle 2} } \, 
      {\lambda_{jqn}^{\prime} \over g_{\scriptscriptstyle 2} } \; 
\eeq
\end{itemize}


\section{Some illustration on the validity of the approximate formulas through
 numerical calculations}

In order to see how well our approximated formulas of the 1-loop neutrino mass 
corrections work, we present here some of the numerical neutrino mass
values from the approximated formulas and compare them verses those from the
exact expressions of the corresponding neutrino mass terms. We have a 
disclaimer to pronounce first. What we do here is not a numerical studies of
phenomenological viable scenarios of neutrino masses generation within the 
model. We make no attempt to choose parameters to fit any neutrino oscillation
data. Rather, we are choosing simple and quite arbitrary input parameters, 
only to check and give an idea on the validity of out analytical results. The
practice also helps to illustrate some theoretical issues behind the formulas.
We choose a set of convenient input parameters and compute and list 
results from nine of the long list of neutrino mass terms. While the results
do give some idea on the relative strength of the various terms, the readers
should be warned that this is only a consequence of a specific choice of
inputs, which is in no sense generic or particularly phenomenologically
interesting.

Our choice of input parameters is as follows. We take the SUSY mass as
around the scale of  100 GeV . In the exact results calculations, however,
we have to split the masses of different superpartners to avoid unwanted
special cancellations. We will clarify on the latter issue below. We choose
input values that turned up mass eigenvalues for the SUSY particles in
the hundreds of GeV scale, details of which is really not interesting. The
value of  $\tan \beta$ is set at 3. The parameters responsible for the lepton
number violating effect are simply taken to be the same numerically, at
a value of $10^{-4}$. Explicitly, 
$\lambda_{ijk}=\lambda^{\prime}_{ijk}=10^{-4}$, 
$\mu_i=10^{-4}\,\mbox{GeV}$, $B_i=10^{-4}\,\mbox{GeV}^2$. 
The neutrino mass results are presented in table 1, in which we show only 
contributions to the (3,3) elements of the effective (SM) neutrino mass matrix.
The upper part of the table corresponds to the results from the approximated formulas 
while the lower part to those from the corresponding exact expressions. In each part, 
the first line corresponds to tems in Eqs.(42,43,B2), the second one to 
Eqs.(B3,B6,B15) and the third one to Eqs.(B22,48, C1). 

As one can see from the table, the difference between the two results for any
specific contribution is within an order of magnitude. One could not really expect
a better agreement than this. In fact, as discussed above and in some related
earlier studies\cite{as46,as7}, the structure of the class of 1-loop diagrams are
such that there is a GIM-type unitarity cancellation involved in the sum over
mass eigenstates. Say, if all the mass eigenstate fermions of the same quantum
number are degenerate, the sum over the set of fermion mass eigenstates in a
neutrino mass term will be proportional to the mass matrix entry that a naive
look at the Feynman diagram will suggest. In most cases, that is vanishing.
Similarly, when the set of the mass eigenstate scalars involved in a certainly 
diagram is mass degenerate, the sum over the set of states gives a vanishing
result due to unitarity of the diagonalizing matrix. Take Eq.(B6) as an illustrative
explicit example, the exact expression of the neutrino mass contribution is
proportional to 
\[
-\mu_j \; \frac{C_4}{M_s} \; 
\frac{y_{\!\scriptscriptstyle e_k}}{g_{\scriptscriptstyle 2}} \,
\mbox{\boldmath $V$}^*_{\!\!(k+2)n}  \, {M}_{\!\scriptscriptstyle \chi^{\mbox{-}}_{n}}  \,
\mbox{\boldmath $U$}_{\!1n}\, D^{l}_{\!(k+2)m} \, D^{l^*}_{\!(i+2)m} \;
B_0\!(p^2, {M}_{\!\scriptscriptstyle \chi^{\mbox{-}}_{n}}^2, M_{\tilde{\ell}_m}^2) \; .
\]
In case of mass degeneracy, one can factor out a fermion summation 
\[ \sum_n 
\mbox{\boldmath $V$}^*_{\!\!(k+2)n}  \, {M}_{\!\scriptscriptstyle \chi^{\mbox{-}}_{n}}  \,
\mbox{\boldmath $U$}_{\!1n}
\]
and a scalar summation
\[ \sum_m
 D^{l}_{\!(k+2)m} \, D^{l^*}_{\!(i+2)m} \;.
\]
The former is nothing but the vanishing $(k+2,1)$ entry of the charged fermion
mass matrix, while the latter is zero by unitarity of the matrix $D^l$ (for $i\ne k$).
As discussed in Refs.\cite{as46,as7}, the lack of mass degeneracy leads to first
order violation of such unitarity cancellations, which explains the nonvanishing
results. It also explains the not better than order of magnitude agreement 
between our exact results, obtained really summing over all the
contributions from the different mass eigenstates, and that from the approximate
formulas, which only extract the analytical form of the largest term within such
summations.

With the above explanation, we see that our approximate formulas do work as
well as they are to be expected. We emphasize again that the approximate 
formulas mainly serve the purpose of illustrating the role of the lepton
number violating parameters in each of the neutrino mass contribution term.

\begin{table}[h]  
\noindent
Table I. Some numerical results from the chosen neutrino mass terms. (See text 
of Appendix E).
\begin{tabular}{ccc}
\hline
&& \\
& $\mbox{approximated formulae (eV)}$& \\
&& \\
\hline
&& \\
$-3.85\times 10^{-9}$ & $-9.35 \times 10^{-9}$ & $6.38 \times 10^{-8}$\\
&& \\
$1.78 \times 10^{-10}$ & $-4.38 \times 10^{-8}$ & $-2.47 \times 10^{-9}$ \\
&& \\
$1.00\times 10^{-8}$ & $3.31 \times 10^{-7}$ & $4.46 \times 10^{-3}$ \\
&& \\
\hline \hline
&& \\
& $\mbox{exact formulae (eV)}$& \\
&&\\ \hline 
&& \\
$-2.61\times 10^{-9}$ & $-3.45 \times 10^{-8}$ & $1.45\times 10^{-8}$\\
&& \\
$5.63 \times 10^{-10}$ & $-1.73 \times 10^{-8}$ & $-3.94 \times 10^{-9}$ \\
&& \\
$-3.23\times 10^{-9}$ & $6.90 \times 10^{-7}$ & $3.26 \times 10^{-3}$ \\
&& \\
 \hline
\end{tabular}

\end{table}



\begin{thebibliography}{99}
\bibitem{pd}
L.E. Ib\'a\~nez and G.G. Ross,
Nucl. Phys. {\bf B368}, 3 (1992).
\bibitem{nu}
For recent reviews, see M.C.~Gonzalez-Garcia and Y. Nir, hep-ph/0202058;
J.N.~Bahcall, M.C.~Gonzalez-Garcia, and C.~Pena-Garay, hep-ph/0204314.
\bibitem{as12}
O.C.W. Kong, 
{\it Thirty Years of SUSY}, Nucl. Phys. B (Proc. Suppl.) {\bf 101}, 421 (2001).
\bibitem{as8}
O.C.W. Kong, IPAS-HEP-k008, hep-ph/0205205, {\it to be published in} Int. J. Mod. Phys. A (2003).
\bibitem{as5}
O.C.W. Kong,  JHEP {\bf 0009}, {\it 037} (2000).
\bibitem{ru1}
M. Bisset, O.C.W. Kong, C. Macesanu, and L.H. Orr,
Phys. Lett. {\bf B430}, 274 (1998).
\bibitem{ru2}
M. Bisset, O.C.W. Kong, C. Macesanu, and L.H. Orr,
Phys. Rev. {\bf D62}, {\it 035001} (2000).
\bibitem{rpv}  See, for example,
G. Bhattacharyya, Nucl. Phys. B (Proc. Suppl.) {\bf 52A}, 83 (1997); 
H.~Dreiner, {\it Perspectives on Supersymmetry} (ed. G. Kane), p.462, (World Scientific 1999). 
\bibitem{skk}
See, for example, E.J.~Chun, S.K.~Kang, C.W.~Kim, and U.W.~Lee, Nucl. Phys. {\bf B544}, 89 (1999); 
S.Y.~Choi, E.J.~Chun, S.K.~Kang, and J.S.~Lee, Phys. Rev. {\bf D60}, 075002 (1999);
M.~Hirsh, M.A.~Diaz, W.~Porod, J.C.~Romao and W.F.~Valle, Phys. Rev. {\bf D62}, 113008 (2000). 
We have no intention to give a complete list of references, interested
readers may check out reference lists of some of the papers cited here.
However, we would like mention also a couple of more noteworthy
earlier works. These include, R.~Hempfling, Nucl.Phys. {\bf B478}, 3 (1996);
B.~de~Carlos and P.~L.~White,  Phys. Rev. {\bf D54},  3427 (1996);
H.~P.~Nilles and N.~Polonsky, Nucl.Phys. {\bf B484}, 33 (1997).
\bibitem{ru6}
O.C.W. Kong, Mod. Phys. Lett. {\bf A14},  903 (1999).
\bibitem{DL}
S. Davidson and M. Losada, JHEP {\bf 0005}, {\it 021} (2000);
Phys. Rev. {\bf D65},  {\it 075025} (2002).
\bibitem{kias}
E.J. Chun and S.K. Kang, Phys. Rev. {\bf D61},  {\it 075012} (2000).
\bibitem{as1}
K. Cheung and O.C.W. Kong, Phys. Rev. {\bf D61}, {\it 113012} (2000). 
\bibitem{neutra} 
S.~Kiyoura, M.M.~Nojiri, D.M.~Pierce, and Y.~Yamada,
Phys. Rev. {\bf D58},  {\it 075002} (1998). See also
D. Pierce and A. Papadopoulos,  Phys. Rev. {\bf D50},  565 (1994) for more background.
\bibitem{nrge}
See, for example, M. Frigerio and A. Smirnov, JHEP {\bf 02}, 004 (2003) 
and references therein.
\bibitem{as4-7}
Ref.\cite{as5} identifies pieces of RPV contributions to
squark and slepton masses often overlooked in the literature and first
address their phenomenological implications. Some more detailed studies
of the related stringent constraints are presented afterwards for
neutron electric dipole moment\cite{as46} and $\mu \to e\,\gamma$\cite{as7}. 
\bibitem{as46}
Y.-Y.~Keum and O.C.W.~Kong, Phys. Rev. Lett. {\bf 86}, 393 (2001);
Phys. Rev. { \bf D63}, {\it 113012} (2001).
\bibitem{as7}
K.~Cheung and O.C.W.~Kong, Phys. Rev. {\bf D64},  {\it 095007} (2001).
\bibitem{rd} 
S.K.~Kang and O.C.W. Kong, {\it work in progress}.
\bibitem{GH}
Y. Grossman and H.E. Haber,  Phys. Rev. {\bf D59}, {\it 093008} (1999); hep-ph/9906310.

\end{thebibliography}
\end{document}